\documentstyle[aasms4]{article}
\newcommand{\beq}{\begin{equation}}
\newcommand{\eeq}{\end{equation}}

\received{}
\accepted{}
\slugcomment{ Submitted to Astron. J.}

\lefthead{Benedict et al.}
\righthead{Star Clusters in NGC 4314}

\begin{document}
\def\vi{\hbox{$V\!-\!I$}}
\def\kms{km\thinspace s$^{-1}$\ }
\def\Ha{H$\alpha$}
\def\etal{{\it et al.~}}

\title{NGC 4314. IV.\ Photometry of Star Clusters with {\it Hubble Space Telescope} - History of Star Formation in the Vicinity of a \\Nuclear Ring\footnote{Based on observations made with
the NASA/ESA Hubble Space Telescope, obtained at the Space Telescope
Science Institute, which is operated by the
Association of Universities for Research in Astronomy, Inc., under NASA
contract NAS5-26555}}

\author{ G.\ Fritz Benedict\altaffilmark{2}, D. Andrew Howell\altaffilmark{3},\altaffilmark{7},
Inger J\o rgensen\altaffilmark{4}, 
Jeffrey D. P. Kenney\altaffilmark{5} , and Beverly J. Smith\altaffilmark{6}}
\altaffiltext{2}{McDonald Observatory, University of Texas, Austin, TX 78712}
\altaffiltext{3}{Astronomy Dept., University of Texas, Austin, TX 78712}
\altaffiltext{4}{Gemini North Observatory, Hilo, HI, 96720}
\altaffiltext{5}{Astronomy Dept., Yale University, New Haven, CT, 06520}
\altaffiltext{6}{Dept. of Physics and Astronomy, East Tennessee State University, Johnson City, TN, 37614}
\altaffiltext{7}{now at Lawrence Berkeley Lab, University of California, Berkeley, CA 94720}
% Notice that each of these authors has alternate affiliations, which
% are identified by the \altaffilmark after each name.  The actual alternate
% affiliation information is typeset in footnotes at the bottom of the
% first page, and the text itself is specified in \altaffiltext commands.
% There is a separate \altaffiltext for each alternate affiliation
% indicated above.

% The abstract environment prints out the receipt and acceptance dates
% if they are relevant for the journal style.  For the aasms style, they
% will print out as horizontal rules for the editorial staff to type
% on, so long as the author does not include \received and \accepted
% commands.  This should not be done, since \received and \accepted dates
% are not known to the author.

\begin{abstract}

Using {\it HST} WFPC2 images, we have obtained 
$U, B, V, I,$ and H$\alpha$ photometry for 76
star clusters in the nuclear star-forming ring 
of the barred spiral galaxy NGC 4314. These clusters are likely associated with an inner Inner Lindblad 
Resonance, or IILR.
The blue colors and \Ha~emission for
most of these clusters imply very young ages of 
1-15 Myr. Age estimates based on several reddening-free parameters indicate that the present epoch of star formation has lasted at least 30 Myr. By estimating the masses of stars in the clusters and comparing with the \Ha~luminosity, we conclude that
a significant fraction of ongoing star formation in the nuclear ring of NGC 4314 occurs in clusters. The cluster masses identify these as young open clusters, not young globular clusters. Further out in the galaxy, just exterior
to the ring of young stars, previous ground-based observations
revealed two 
symmetric stellar spiral arms which may be associated with an outer Inner Lindblad Resonance, or OILR. 
With our {\it HST} data, we have revealed part of this
structure and its colors in more detail.
The spiral arm colors are consistent
with stellar ages between 40 and 200 Myr. 
The age difference between the inner ring of young
stars (IILR) and the larger oval-like feature containing the blue arms (OILR)
supports an interpretation of the morphology of the nuclear region of NGC 4314 that requires a reservoir of gas that becomes more compact over time. We speculate that as the gas distribution becomes more centrally concentrated, it
interacts with these two resonances. Each resonance triggers star formation, resulting in two distinct epochs of star formation. 

\end{abstract}

\section{Introduction}

The primary goal for this paper is to determine the history of star formation associated with the nuclear ring of NGC 4314. We seek details of the history of recent and past star formation to choose among a number of possible star formation processes. Buta \& Combes (1996) reviewed progress towards an understanding of rings in galaxies. They describes rings as a natural consequence of barred galaxy dynamics and that rings are often the only location within a galaxy experiencing active star formation. A complete understanding of star formation in rings will only come from detailed examination of that ongoing process. NGC 4314 is one of the nearest examples of a galaxy hosting a nuclear ring of active star formation. For NGC 4314 this process take place in a galaxy otherwise devoid of star formation, against the smooth, uncomplicated backdrop of an early barred spiral galaxy. To determine a star formation history of NGC 4314 we analyze UBVI\Ha~ photometry of clusters in the nuclear ring. We obtain evidence of past but recent star formation from surface photometry of regions just exterior to the nuclear ring.

To summarize past work on this galaxy, we note that NGC
4314 exhibits
evidence for many features associated with resonances. It has a large-scale stellar
bar of diameter 130$^{\prime \prime}$ (8.3 kpc) and a prominent
circumnuclear ring of  star formation of diameter 10$^{\prime
\prime}$  (640 pc) that is visible in \Ha~(Pogge  1989),
radio continuum (Garcia-Barreto et al.\ 1991),
and  optical color maps (Benedict et al.\ 1992, hereafter P1). CO
(1-0) mapping  of NGC 4314 (Combes et al.\ 1992) at 5$^{\prime
\prime}$  resolution revealed the presence of a molecular ring whose mean radius is
slightly smaller than the ring of star formation. Benedict et al.
(1996, =P3) mapped this region in CO with 2\farcs2 spatial and 13 \kms velocity resolution
and detected molecular gas flowing radially into the ring. The galaxy morphology and kinematics (P1, P3) both strongly suggest that the nuclear ring is associated with an Inner Lindblad Resonance (ILR).

Outside of the stellar and gas rings, a blue elliptical feature of diameter
20-25$^{\prime 
\prime}$  (1 kpc)  is seen in optical and IR color maps, which may
correspond to a ring of relatively young but non-ionizing stars
(P1). This structure was identified with an Outer Inner Lindblad Resonance (OILR) in P3. There are 2 symmetric spiral arms within this elliptical region, which can be seen on the  unsharp masked
optical image shown in figure 1 of P1 and in the optical-IR color map presented by \cite{Woz95}. This elliptical region is elongated 
perpendicular to the primary bar, suggesting that it is partially composed
of stars on x$_2$ orbits that may be associated with an ILR (Athanassoula 1992). Combes et al.\ (1992) contend that in this
galaxy a ring of molecular gas is  propagating inwards. This is in
contrast with the evolutionary scenario  proposed by Kenney et
al.\ (1993) for strong starbursts, in which star formation
devours gas most rapidly in the center, ultimately forming a 
ring of gas near the ILR. 

\begin{figure}
\epsscale{1.0}
\plotone{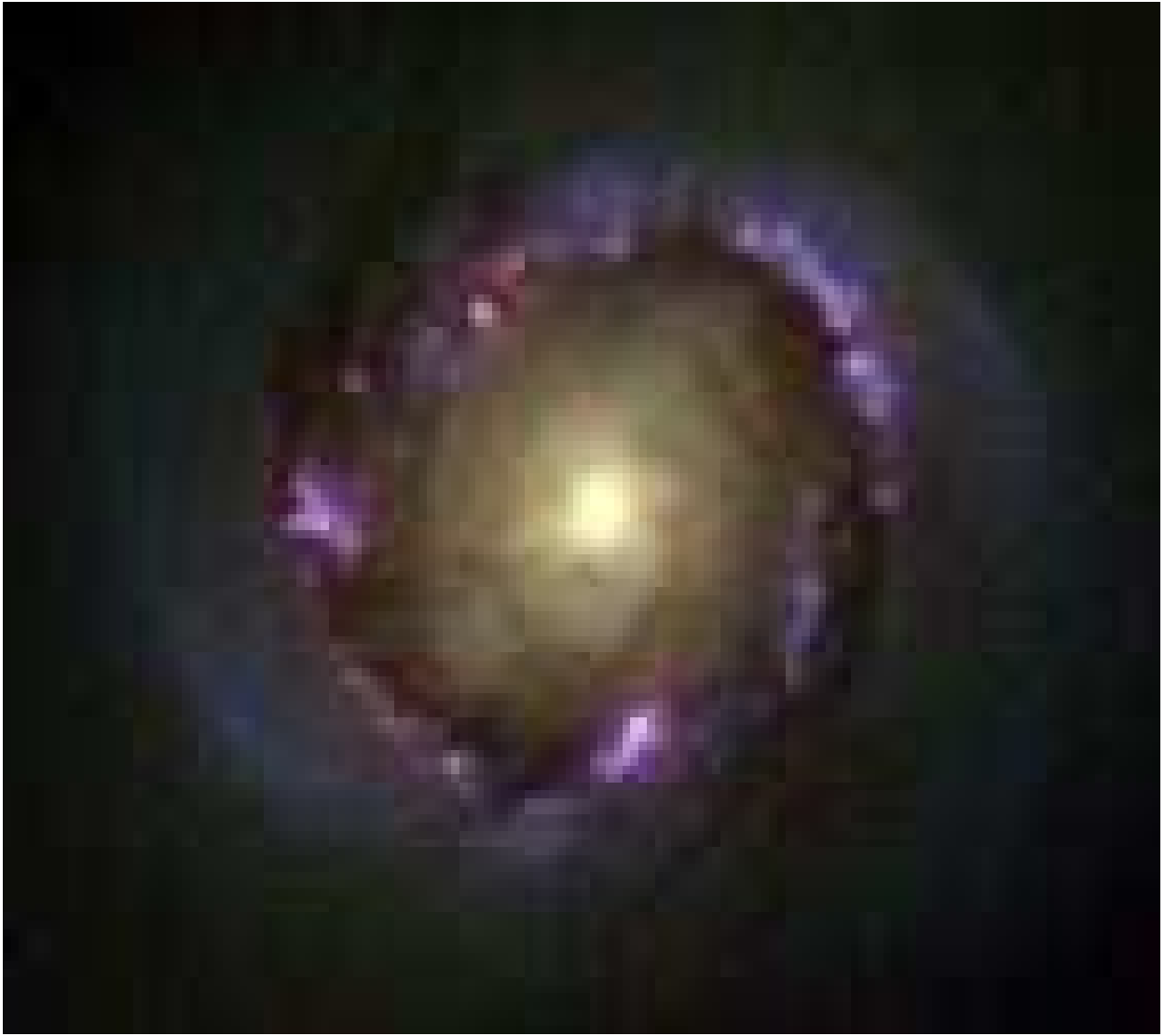}
\caption{True color image of the nuclear ring of NCG 4314, North at top, East to left. U + B provided the blue channel, V the green, and red was a combination of I + \Ha. Note the purple hue of the strongest H {\sc ii} regions. Note the blue arms to SE and NW, exterior to the ring of presently active star formation. The field of view is 25\farcs7$\times$23\farcs0.} 
\label{Benedict.fig1}
\end{figure}

Much evidence of the processes that trigger the conversion of gas into stars can be seen in the active nuclear ring of NGC 4314. The color image of the nuclear ring in the barred spiral galaxy NGC 4314 shown in Figure~\ref{Benedict.fig1} serves to summarize the motivations for the present study. We are presented with a narrow ring of active star formation, a distribution of dust and CO (P3) delineating the raw material for that formation, and possible evidence of past star formation -  the blue arms just exterior to the ring to the SE and NW. Understanding those blue arms is our secondary goal.
In order to do this, we model and subtract the underlying bulge and bar
stellar distributions, which are composed of presumably older stars. Finally, we wished to exploit WFPC-2 resolution to re-confirm the nuclear bar first detected in lower-quality WFPC-1 data (\cite{Ben93}, P2), now confirmed with NICMOS observations (Ann, 2001).

We describe our observations 
in Section 2 
and discuss data reduction and calibration in Section 3. We present our surface photometry in Section 4, where we confirm the nuclear bar and obtain colors corrected for the background light from the galaxy for regions just exterior to the nuclear ring, including the blue spiral arms.  In Section 5 we look for patterns in the spatial distribution of the star clusters in the nuclear ring and in the blue arms. In Section 6 we interpret the photometry of the clusters associated with the nuclear ring to obtain cluster reddening, absolute magnitudes, ages, and masses. We briefly analyze the \Ha~morphology. In Section 7 we compare ages obtained for the stars comprising the blue arms with those derived for the nuclear ring clusters. In Section 8 we discuss our observations in the context of star formation rates and processes. Throughout this paper we  compare our results to those on  the equally well-studied nuclear ring galaxies, NGC 1326 (\cite{But00}) and NGC 1512 (\nocite{Mao01}Maoz \etal 2001). We summarize our results in Section 9. Some general properties of NGC 4314 relevant to this paper are provided in Table 1. We adopt a distance of 13.1 Mpc, derived from the CO systemic velocity obtained in P2 and H$_0$=75 \kms Mpc$^{-1}$ (\cite{Mad99}).

\section{Observations}
The observations were acquired using the {\it Hubble Space Telescope} WFPC-2 camera on 29 December 1995. During five orbits we obtained aggregate exposures of  3700s, F336W (U); 2500s, F439W (B); 600s, F569W (V); 600s, F814W (I); and 600s, F658N (\Ha). Note that F658N ($\lambda = 659.0$ nm) was used rather than the standard \Ha~filter, F656N, because the redshift of NGC 4314 would have removed \Ha~from the 28.5 \AA~bandpass. This paper only discusses data from the higher-resolution Planetary Camera (PC) chip on WFPC-2. The PC has a pixel size of 0\farcs046 and a field of view 36\farcs8$\times$36\farcs8.

\setcounter{footnote}{0}
\section{Data Reduction and Calibration}
\subsection{Images}
All data reduction was done using 
IRAF\footnote{IRAF is distributed by the National Optical Astronomy Observatories,
    which are operated by the Association of Universities for Research
    in Astronomy, Inc., under cooperative agreement with the National
    Science Foundation.}. First, all images were shifted to the same coordinate system using the task 
{\tt imshift}.  The individual images for each filter were then 
co-added using the option {\tt crreject} with the task {\tt combine}
to remove cosmic rays.
Images were scaled by exposure time. Conversion of instrumental magnitudes to Landolt $UBVI$ magnitudes was
done using the calibration data presented in Holtzman et al. (1995b = 
H95b).

The color image in Figure~\ref{Benedict.fig1} was produced by Zolt Levay at the Space Telescope Science Institute. For this image U + B provided the blue channel, V the green, and red was a combination of I + \Ha.

\subsection{Nuclear Ring Cluster Aperture Photometry} \label{NRCAP}
A master image, made by combining data from all wavelengths, was 
used for nuclear ring cluster (NRC) identification.  {\tt Daofind} was run on the master
image. Those 186 objects that were detected
at the 10 sigma level were identified as candidate clusters.
The coordinates of these objects were used as input to {\tt daophot},
which was used to generate the instrumental photometry.

The size (full width half-maximum) of each source is one of the parameters returned by {\tt daophot}. Very few clusters had a FWHM greater than that of a stellar source (FWHM $>$ 2.7 pixels = 0\farcs12 = 7.5pc). All but one of those `resolved' were in crowded fields. We concluded that, except for one cluster, resolution was spurious due to crowding. In support of this assertion
we point out that the Pleiades are about 2 pc across,
as is R136 in the LMC (Hunter et al. 1997). The median cluster diameter in NGC 4449 (\cite{Gel01}) is 5pc and $\sim$3pc in NGC 1569 (\cite{Hun00}). The one resolved cluster is NRC 42, discussed in Section \ref{NOIC}.

Due to this crowding of clusters, point spread function (PSF) fitting
could not be used, so aperture photometry was necessary.  Small apertures
were necessary to reduce errors due to the crowded nature of the ring.
An aperture radius of 0\farcs2 (4.4 pixels for the PC) was chosen.
The median sky value was determined from an annulus of width 4 pixels starting
5 pixels from the center of the cluster.

\subsubsection{Photometric Calibration} \label{PHOT}
Conversion of instrumental magnitudes to Landolt $UBVI$ magnitudes was
done using the calibration data presented in H95b.
Because Holtzman et al. used an aperture radius of 0.5 arcsec, an 
aperture correction was necessary to correct the 0\farcs2 radius to 0\farcs5 radius.  
Aperture corrections of 0.11 magnitudes for filters F336W, F439W, and 
F814W, and 0.12 mag for F569W were derived from the encircled energy curves 
presented in Holtzman et al. (1995a = H95a).

The focus position of {\it HST} is known to drift over time.  The effect of
this drift is to change the photometric zero points. The zero points used here
have been corrected (Suchkov \& Casertano 1997) to take into account focus drift.  Corrections of
-0.04, -0.02, and -0.03  mag were applied to the zero points presented in 
H95a for F439W, F555W, and F814W respectively.

The impact of charge transfer efficiency (CTE) on these data is thought to
be negligible. The observations were taken on December 29, 1995.  This is after
 the temperature of WFPC2 was changed to --88$\arcdeg$C, which reduced the
 magnitude of the effect, but before recent increases in CTE effects
 (Whitmore et al. 1999b). 
In addition, the presence of significant background ($>250$ $ e^-$) in these
images makes a correction for CTE unnecessary (H95a).

We obtained a continuum-free H$\alpha$ image of NGC 4314 by 
subtracting the I image, after converting both images to 
units of erg s$^{-1}$ cm$^{-2}$ \AA$^{-1}$, using the calibration
constants from the WFPC2 Data Handbook.  Generally for 
H$\alpha$ continuum subtraction, an R band image or a
narrowband image is used instead of I (e.g., \cite{Ken89}).  To test how much uncertainty is introduced
in our flux values by substituting the I band image,
we created
a second continuum-subtracted image using the V band image.
We found a very similar morphology, and the total flux
density agrees within 2$\%$ of the values obtained with the I band image.  Our final map is an average of the two
continuum-subtracted images, roughly an (\Ha + continuum) - 'R' image. The total H$\alpha$ flux that we obtained
for the galaxy is 4.3 $\times$ 10$^{-13}$ 
erg s$^{-1}$ cm$^{-2}$,
compared to 1.0 $\times$ 10$^{-13}$
erg s$^{-1}$ cm$^{-2}$ quoted by Gonzales Delgado
et al. (1997) based on ground-based data. 

In partial explanation for the discrepancy it is likely that the \Ha~emission line was at the edge of the 5.3 nm bandpass of their filter (centered on 656.3 nm).
Also, our F658N measures include flux from the [N{\sc ii}] lines redshifted to  657.0 and 660.5 nm. The first of these is on the short wing of the F658N filter at $\sim$40\% transmission. The second is at $\sim$80\% transmission. We estimate (from integral field spectroscopy obtained with TIGER/OASIS on the CFHT) that the \Ha/[N{\sc ii}] ratio in the vicinity of the cluster complex involving NRC 53, 54, and 57 is $\sim$50\%. Assuming this level of [N{\sc ii}] contamination in our filter for all sources it is likely
that our true \Ha ~flux is $\sim$ 2 $\times$ 10$^{-13}$
erg s$^{-1}$ cm$^{-2}$. The total \Ha+[N{\sc ii}] luminosity of the NGC 4314 nuclear ring is $9\times10^{39}$ erg s$^{-1}$, compared to $6\times10^{39}$ erg s$^{-1}$ for NGC 1512 (Maoz \etal 2001). Both these galaxies have a total \Ha+[N{\sc ii}] luminosity
an order of magnitude less than that of the nuclear ring in NGC 1326, where Buta \etal (2000) find L(\Ha)= $1.2\times10^{41}$ erg s$^{-1}$.

\subsubsection{NRC Photometry Internal Errors} \label{CUT}
Because the systematic errors inherent in converting the instrumental 
magnitudes to the Landolt system are particularly hard to characterize,
only the statistical errors are presented here.  Systematic errors should
be small for all bands except U, which as H95b points out, may have
errors of 0.1 mag or greater.  However, since we are concerned mainly with
relative colors, systematic effects should not affect our results.
With our primary goal the search for age differences, we reduced the effects of faint clusters and/or poorly determined sky
values due to crowding of the field by selecting only 
NRC with statistical $V-I$ and $U-B$ color errors less than 0.15 mag.  Out of an initial sample of 186 NRC, 76 met these 
criteria, all with V $\le$ 22.90. These NRC are identified in a finder chart, Figure ~\ref{Benedict.fig2}. 

\subsubsection{Correction for Galactic Extinction}
Despite NGC 4314 having a high galactic latitude (+83\arcdeg), the extinction map of \cite{Sch98} yields $E_{B-V}=0.024\pm0.004$, 
indicating an extinction due to our Galaxy of A$_V$ = 0.083. Assuming a
Savage \& Mathis (1979) \beq
R_V= {A_V \over E_{\bv}} = 3.1
\eeq
extinction curve 
we correct our measured NRC colors: (\ub)$_O$= \ub - 0.028, (\bv)$_O$= \bv - 0.024, and (\vi)$_O$ = \vi - 0.035. These corrected colors are used for the remainder of this paper. We provide RA and Dec positions (relative to the galaxy center), V$_O$, \ub$_O$, \bv$_O$,  and \vi$_O$ for the NRC in Table 2. We discuss the \Ha + [N{\sc ii}] fluxes in section 6.3 below.

\section{Surface Photometry - Modeling the Nuclear Region with Nested Ellipses}
We model the light distribution within the PC using the nested ellipse approach described in J\o rgensen et al. (1992). The data modeled include only contributions from the bulge and primary stellar bar. The regions of the nuclear ring and blue arms exterior to the ring were excluded from the fit. Figure~\ref{Benedict.fig3} provides the results of this modeling as the variation with radius of V surface magnitude ($\mu_{V}$), \vi~surface color, ellipse position angle, eccentricity, and ellipse center. 
In Figure~\ref{Benedict.fig4} we show the Fourier coefficients $c_3, c_4, s_3$, and $s_4$ (e.g., $c_4cos4\theta$). All parameters except $\mu_{V}$ and \vi~were determined from the I-band data.

Excluding the blue arms (seen in Figure~\ref{Benedict.fig1}) exterior to the ring from the ellipse modeling permits us to obtain their intrinsic colors. We obtain this information from the residual maps, by subtracting the fitted ellipses from the original data.  The excess signals above the model in each bandpass is used to produce color indices. Figure~\ref{Benedict.fig5} presents the sum of all the residual maps (U, B, V, and I). The sum yields the least noisy and most detailed map of the dust and star cluster distributions. It is obvious that most of the dust (hence, gas: see P3 section 3.2.2) lies interior to the newly formed star clusters. 

\subsection{The Nuclear Bar}
Inside the H\,{\sc ii} region ring, a nuclear stellar bar of
diameter $\sim 8^{\prime 
\prime}$  (480  pc) was seen in {\it HST}
WFPC-1 I band data (P2). The nuclear bar lies within the IILR (Binney \& Tremaine 1987).
Ann (2001) used an HST H-band image to reveal a nuclear bar by subtracting off an r$^{1/4}$ bulge component. At the radii where
the luminosity from the nuclear bar is dominant, from about
3 arcsec to 6 arcsec (Ann 2001), our new HST/WFPC2 data show (Figure~\ref{Benedict.fig3}b) a 
significantly different position angle (PA=140\arcdeg) compared to
the position angles for smaller radii. We take PA=140\arcdeg~ to be the  position angle of the bar. We also see a weak maximum in the 
ellipticity profile, though the ellipticity does not reach as high a 
value as found by Ann. This difference is most likely due to the 
difference in wavelengths for the observations.  Of particular interest is the
strength of the cos(4$\theta$) term in the Fourier expansion of the isophotal deviation
from a pure ellipse. The $c_4$ coefficient 
is positive for the 3 arcsec to 6 arcsec  radius interval, indicating that the isophotes are 
boxy (e.g. Lauer 1985, Mihos \etal 1995). Whether a bar is detected as purely disky or purely boxy depends on the
surrounding luminosity distribution. In the case of NGC 4314, there are
dust lanes in and near the center of the galaxy (see Figure 5), reducing the amplitude of the $c_4$ coefficient to a 'weakly boxy' indicator.

The intricate dust distribution revealed in Figure~\ref{Benedict.fig5} shows lanes extending to the center of the galaxy. This may suggest gas inflowing to within a few pc of the galaxy nucleus. The nuclear bar may be partly responsible for driving this gas inwards, although the complex dust lane pattern suggests that 
other processes may be dominant (\nocite{Shl01}Shlosman \& Heller 2001).
The nucleus of NGC 4314 is not strongly active; there are no broad line components detected, although its nuclear emission line ratios give it a weak LINER classification (Ho et al. 1995).

\subsection{Surface Photometry of the Blue Arms} \label{SPBA}
Our approach for obtaining photometry of the stars comprising the blue arms is predicated on two assumptions. The first is that galaxy shape is defined by stellar dynamics. The shape of the bulge and primary stellar bar combination
is determined by the dynamics of two old stellar populations. The second assumption is that the blue arms represent an additional, dynamically distinct, population.

Twelve apertures (radius of 0\farcs2, same as for the cluster photometry) were placed along the blue arms just outside
of the nuclear ring, as shown in Figure~\ref{Benedict.fig5}. After subtracting the flux predicted from the nested ellipse model,
residual fluxes inside these apertures were measured and converted to 
$U, B, V, I$ surface magnitudes and \bv, \ub, and \vi ~colors for the purpose of comparison to the NRC. Again, correcting for extinction within our Galaxy, we provide surface magnitudes and colors $\mu_{V_O}$, $\mu_{(U-B)_O}$, $\mu_{(B-V)_O}$, and $\mu_{(V-I)_O}$ for these locations in Table 3, where $\mu_x$ denotes surface photometry per square arcsec in bandpass or color index $x$.

\section{The Spatial Distribution of the NRC and the Blue Arms} 
The morphology of the nuclear region of NGC 4314 stands as the final arbiter between various dynamical models that attempt explanation of past, present, and future distributions of stars and gas in this nuclear ring. Two recent studies bearing on this nuclear morphology are an SPH model of the galaxy itself (Ann 2001) and the analytic models of Byrd et al. (1998). Ann attempts to generate nuclear bar-driven spirals in the gas distribution. Byrd et al. explore the general properties of resonance rings. Hence, ring shape is a consequence of stellar dynamics and might provide evidence for a particular formation process, particularly when combined with age estimates (sections 6.4, 6.5, and 7).

We would like to know the radial and azimuthal distribution of the NRC and the blue arms with respect to the galaxy center. We have compared the distribution of the NRC and the shapes of the blue arm with three simple model
distributions: circular, elliptical, or spiral. 

We first assume that the clusters are coplanar with the principal plane of the galaxy with an intrinsically circular distribution. The elliptical distribution seen in Figure~\ref{Benedict.fig2} would then be a consequence of the inclination of the galaxy. To deproject the nuclear ring, an ellipse was fit to the cluster positions
in the ring.
There are five parameters of interest - x and y coordinates for the center of 
the ellipse, semi-major and semi-minor axes, a and b, and the position angle of 
the ellipse in the plane of the sky, $\theta$.  The input data consisted of 
the positions (in pixel coordinates) of {\it all} the clusters detected in all 
passbands (U,B,V,I, H-alpha) by the {\tt daofind} routine with parameters set 
as reported earlier.  Clusters not obviously associated with the ring (either closer to or further away from the galaxy center) were removed, leaving 153 objects.  

To fit the NRC distribution we employed a genetic algorithm, a {\tt FORTRAN} program \footnote{http://www.hao.ucar.edu/public/research/si/pikaia/pikaia.html} written specifically 
to do ellipse fitting to discrete points (Charbonneau 1995). (The STSDAS routine {\tt ellipse} fits ellipses to isophotes, not to a sparse distribution like that of our clusters.) 
Genetic algorithms work by encoding parameters of interest as strings representing
members of a single species.  Different solutions are then subjected to computational 
evolution by natural selection (the best solutions from each iteration
``breed" new child solutions), mutations (individual bits in the strings 
are randomly changed), and crossover (parts of strings are swapped). Ultimately the `best' solution is encoded in the surviving string. One of the major strengths of this approach is the very wide parameter space that can be searched for 'best' solutions. Unlike many non-linear least-squares approaches, no initial guess is required. However, no uncertainty estimates are produced, a drawback of the genetic algorithm approach.

The parameters of the ellipse fit determined by the genetic algorithm
are as follows.  
The center of the ellipse was determined to be
within 1.5 pixels (0.07 arcsec) of the galaxy 
nucleus. The semimajor axis of the ellipse was found to be  
7\farcs4 with ellipticity $\epsilon$=0.28 and position angle PA= 137\arcdeg. If the nuclear 
ring is inherently circular, then an inclination of 44\arcdeg~ 
is derived. Note that the inclination and position angles differ from values determined in P1 using the outer isophotes of
the galaxy ($i=23 \pm 8$\arcdeg and PA=$51\arcdeg \pm 8\arcdeg$). The PA also differs from that of the line of nodes, $115\arcdeg \pm 10\arcdeg$, determined from OVRO CO data (P3).  This may indicate that the nuclear ring is inclined 
relative to the plane of the galaxy or is not inherently circular. For example, Buta et al. (1999) show that the nuclear ring of ESO 565-11 is intrinsically elliptical, not a circle seen in projection. Ann (2001) deprojects an H-band image of the nuclear ring of NGC 4314, assuming the inclination derived in P1, and finds an intrinsically elliptical distribution with $\epsilon$=0.35. From P1 the primary stellar bar has PA = $146\arcdeg \pm3\arcdeg$.
We note that the PA of the primary bar, the nuclear bar, and the nuclear ring are probably the same within the errors.
The nuclear ring is aligned with the bars, a result confirmed by Ann (2001).

From P1 the semimajor axis of the arcs containing the blue arms was found to be  11\farcs9 with ellipticity $\epsilon$= 0.28 at PA = 48\arcdeg. The apparent ellipticity of the distribution of NRC is identical. On the sky the PA of the distribution of the NRC and the ellipse containing the blue arms differ by 89\arcdeg. This near 90\arcdeg~difference suggests inner and outer Inner Lindblad Resonances (Combes 1988), and supports the IILR/OILR dynamical interpretation from the CO velocity field discussed in P3.

In P1 the spatial distribution of the blue arms was explored.
Our new data (the positions of A1 through A12 on Figure~\ref{Benedict.fig5}) do not change the conclusions reported in P1. Concerning the choice between an elliptical or a spiral shape for the blue arms, a spiral with 
a pitch angle $i = -12\fdg7\pm0\fdg9$ (indicating trailing arms) was weakly favored over an ellipse. 

Our new higher resolution data (the positions in Table 2) suggest that there is no unique spiral pattern that connects the NRC to the blue arms. A single spiral was qualitatively supported by the lower spatial resolution data of P1. The NRC and blue arms might not be simultaneously generated by a single spiral density wave pattern. 

Byrd et al. (1998) discuss resonance rings in galaxies and explain ring misalignment as due to dissipation (cloud-cloud collisions). The OILR and IILR in NGC 4314 show no evidence for dissipative misalignment and match the Byrd et al. model with a weak bar (q$_2$=0.02) and strong clumping (c = 0.03), the top left panel in their figure 2. The morphological consequences of increased dissipation include the narrowing of the gap between the morphological features that trace the OILR and IILR. Increased bar strength
sharpens the ends of the morphological features associated with an OILR. In NGC 4314 the morphological features we associate with the OILR (blue arms) and IILR (distribution of NRC) are nearly touching along the IILR major axis at PA $\sim$148\arcdeg. The morphological feature we associate with the OILR, traced by the A and B components in Figure~\ref{Benedict.fig19}, are quite rounded. We conclude that if the nuclear morphology of NGC 4314 is produced by a nuclear bar, it is a weak bar. Regarding this last point, Shlosman \& Heller (2002) have modeled nuclear bars of varying strengths. The morphology of NGC 4134 most resembles their model 1, a weak nuclear bar.

\section{NRC Photometry}
We wish to establish the sequence of star formation associated with the nuclear ring. The spatial resolution and wavelength coverage of these {\it HST} data permit us to assess cluster ages. Our methodology is similar to that used by
\cite{Whi99a} for the clusters in NGC 4038/39, by Buta et al. (1999) for the nuclear ring in ESO 565-11, by Buta et al. (2000) for the nuclear ring in NGC 1326, and by \cite{Mao01} for the nuclear ring in NGC 1512. We lack the wide wavelength coverage of the NGC 1512 study, so do not fit spectral energy distributions to models.

\subsection{NRC Color-color Diagrams}

We plot in Figure~\ref{Benedict.fig6} the color-color diagrams (\ub)$_O$ ~vs. (\bv)$_O$~ and (\bv)$_O$ ~vs. (\vi)$_O$ for the NRC listed in Table 2. We superpose evolved clusters of various ages from \cite{Lei99}(hereafter SB99), chosen for convenient web access. We plot burst colors, not continuous SFR colors, and chose the $\alpha$ = 2.35 IMF with a truncated upper limit ${\cal M}_{up} =100 ~{\cal M}_{\sun}$ and metallicity, Z=0.020. Note that in the (\bv)$_O$ ~vs. (\vi)$_O$ color-color diagram the youngest SB99 clusters do turn back to redder (\vi)$_O$.

Cluster colors are certainly a function of cluster age, but are also affected by reddening from the interstellar medium local to each cluster. The magnitude of the effect due to $A_V = 1$ is indicated by the reddening line in each color-color diagram. These plots serve to show that neither U-B  vs B-V   nor  B-V  vs V-I are suitable 
for age studies. The reddening and the age variation are approximately 
parallel to one another and cannot be separated. 

As discussed in Whitmore et al. (1999a), the clusters with stronger \Ha~are expected to be too red in \bv ~and too blue in \vi, compared to the model clusters. This is due to emission line contamination in the V bandpass. The `V' we chose, F569W, is somewhat less affected, having less transmission at \Ha~than does the F555W filter.

\subsection{Estimating NRC Reddening with $Q$} \label{QRED}

First introduced by Johnson and Morgan (1953), Q, the reddening-free parameter, is defined
\beq
Q = (U-B)_O - {E_{U-B}\over E_{B-V}}\times(B-V)_O = (U-B)_O-0.72\times(B-V)_O
\eeq
Benedict (1980) showed that Q was applicable to star clusters in external galaxies, specifically that young stars were present all around the nuclear ring in NGC 4314. In Figure~\ref{Benedict.fig7}a we plot Q vs (\bv)$_O$ ~for our clusters and for the SB99 model clusters. The model clusters have ages 0.1 to 100 Myr. We also plot the locus of main sequence stars and a reddening vector produced by $A_V = 1$. The Q values are consistent with the luminosity of each cluster being dominated by OB stars. We estimate the reddening by calculating the distance in (\bv)$_O$ ~between the locus of unreddened model clusters (dotted line at top) and each observed cluster. This provides a color excess, E$_{\bv}$. We then assume (Savage \& Mathis 1979)
R$_V$ = 3.1, because in P1 and P3 we demonstrated a ratio similar to this for a dusty region in the primary stellar bar of NGC 4314, far from new star formation sites. With this ratio we generate the $A_V$ values found in Table 4. For NGC 4314 $<A_V> = 0.6$, half the $<A_V> = 1.2$ found for the nuclear ring of NGC 1326 (\cite{But00}), but identical to NGC 1512 (\nocite{Mao01}Maoz \etal 2001, table 5).

In Figure~\ref{Benedict.fig8} we identify each cluster with a circle proportional to $A_V$ and note that larger $A_V$ values are found to the northeast.  From the CO velocity field presented in P3,
the line of nodes has a position angle of 115\arcdeg $\pm$ 10\arcdeg,
and is redshifted to the northwest.  Assuming that the large-scale 
spiral arms are trailing, the northeast side is the far
side of the ring, consistent with its higher extinction and with
the molecular ring lying inside the ring of star formation.
Comparison of our values of $A_V$ with the CO map in P3 
shows that the ratio of $A_V$ to CO flux is systematically
higher to the northwest, indicating that the star clusters
tend to lie in front of the clouds to the southeast,
and behind the clouds to the northwest. The fact that a far and near side can be inferred from these $A_V$ values lends them credence.

H95b recommend correcting natural magnitudes for
extinction BEFORE using transformations. This is not a problem for
Galactic extinction which comes from an independent measurement (Schlegel, Finkbeiner, \& Davis 1998), but is  a potential problem for the internal extinction values we obtain from Figure~\ref{Benedict.fig7}a. We cannot correct for an extinction that is  unknown until we compare our calibrated Q and $B-V$ values with model cluster colors. We  estimate that the error introduced by not following this order of correction is small. It could be a large effect if we had a mixed population of reddened very old and reddened young
clusters. Evidently the nuclear ring of NGC 4314 does not (Section \ref{QQ3HaA}, below). For our average color excess, $<$E(B-V)$>$=0.2, the maximum difference
in instrumental U-band absorption, A$_U$, is $\Delta A_U = 0.15$ magnitude,
comparing an O6 to a K5 star. As we will see, none of our selected clusters are dominated
by K5 stars. The effect in the other bandpasses is significantly less. See Buta et al. (2000) for a useful discussion of the potential problems associated with color-based age estimates using {\it HST} WFPC-2 photometry.

\subsection{\Ha~Emission Associated with NRC} \label{HaNRC}
We obtained \Ha ~flux measures - using the same 0\farcs4 diameter aperture as for the UBVI photometry - at the location of each cluster identified in
Figure \ref{Benedict.fig2}. Of the seventy-six clusters, 75\% were detected at the 3-$\sigma$ level or higher. Contrast this with NGC 1326  (Buta et al. 2000) where there
is little correlation between \ion H 2 regions and continuum sources (presumably clusters). We list these fluxes in units of $10^{-15}$ erg s$^{-1}$ cm$^{-2}$ (multiplying by the 28.5 \AA~F658N bandpass) and log L$_{H\alpha}$  in Table 5. The average L$_{H\alpha}$ is
2.4$\times10^{37}$ ergs s$^{-1}$, typical for an extragalactic \ion H 2 region
(\cite{Ken89}), and far less than the 5$\times10^{39}$ ergs s$^{-1}$ emitted by 30 Dor in the LMC (Kennicutt and Hodge 1986).

While these \Ha ~fluxes will provide an independent age estimate (see Sections~\ref{CCHaA} and~\ref{QQ3HaA}), they first allow us to test the validity of our derived $A_V$. We see a wide range of \Ha ~fluxes, due either to differences intrinsic to the clusters  (e.g., mass) or variable extinction. To correct for cluster size we form an \Ha ~`equivalent width', $ EW_{H\alpha}$ by subtracting an estimated continuum as follows. 

\beq
EW_{H\alpha} = m_{H\alpha} - R
\eeq
where
\beq
m_{H\alpha} = -18 - 2.5 log(F_{H\alpha})
\eeq
and
\beq
R = 0.55V + 0.45I
\eeq
Again, we plot (Figure~\ref{Benedict.fig7}b) Q vs (\bv)$_O$, where now the symbol size is proportional to $EW_{H\alpha}$, and see uniformity of \Ha ~equivalent width among the younger clusters. Both line and continuum measures suffer about the same extinction,
so $EW_{H\alpha}$ is nearly extinction-independent. 
From the \cite{Sav79} reddening law we determine that the \Ha ~flux 
should experience an absorption, $A_{H\alpha} = 0.806 A_V$ in magnitudes. We correct our m$_{ H\alpha }$ for this extinction and list
the
m$^{c}_{H\alpha}$ in Table 5.
$EW_{H\alpha}$ also indicates a fall-off in \Ha ~flux for
clusters dominated by stars later than B3. To summarize, correcting for extinction improves agreement between age differences from Q and age differences from \Ha. Within the nuclear ring stronger \Ha~is correlated with more negative Q values. 

\subsection{NRC Ages from Color-Color Diagrams and \Ha ~Emission} \label{CCHaA}
With $A_V$ and the \cite{Sav79} reddening law we can estimate absorption in each of our other bandpasses: $A_U$ = 1.66 $A_V$, $A_B$ = 1.32 $A_V$, and $A_I$ = 0.59 $A_V$. We then correct all photometry for absorption intrinsic to NGC 4314 (double O subscripts, e.g. (\bv)$_{OO}$) and arrive at the color-color diagrams presented in Figure~\ref{Benedict.fig9}. The symbol size is proportional to the extinction corrected \Ha ~equivalent width, $ EW_{H\alpha}$. We note the general agreement with the SB99 cluster evolution trajectories in both color-color diagrams, even though only \ub ~and \bv ~(through the Q vs. \bv~plot in Figure~\ref{Benedict.fig7}a) were used to generate the reddening corrections. We also note that the \Ha ~emission and
age are somewhat correlated, with a maximum \Ha ~flux associated with
clusters of age $\sim$3 Myr. Figures~\ref{Benedict.fig7}c and ~\ref{Benedict.fig9} suggest that the
\Ha ~emission from clusters seems to increase up to
an age of about 3 Myr, then decline.

\subsection{Estimating NRC Ages with Several $Q$ Parameters}\label{QQ3HaA}
\cite{Whi99a} have recently estimated ages for clusters in the colliding galaxy pair NGC 4038/4039, The Antennae. They  defined another reddening-free parameter
\beq
Q_3 = (U-B)_O - {E_{U-B}\over E_{V-I}}\times(V-I)_O = (U-B)_O-0.58\times(V-I)_O
\eeq
Each cluster can be plotted on a
Q-Q$_3$ plot and compared with stellar population synthesis
model tracks of cluster evolution (again, we use SB99) to determine its age. 
Figure~\ref{Benedict.fig10} shows this plot for the NRC. 
Comparing with the locations of model clusters of various ages from SB99 we find most of the clusters in our nuclear ring sample are younger than 20 Myr.

With typical errors of 0.10 in Q and Q$_3$, we are unable to search for detailed radial and/or azimuthal age gradients within the ring of active star formation. For crude age resolution we flag those clusters younger than 15 Myr with 'Y' and those older than 15 Myr (15 $<$ age $<$ 40 Myr) with 'O' in Table 5. In the Figure~\ref{Benedict.fig11} Q-Q$_3$ diagram the clusters are represented by symbols whose size is proportional to the extinction-independent \Ha ~equivalent width index, $ EW_{H\alpha}$. Note that ages for clusters with no \Ha~ emission can be estimated from Figure~\ref{Benedict.fig10}. The NRC \Ha ~ equivalent widths support the age differences indicated by their placement on the Q-Q$_3$ diagram. The average $ EW_{H\alpha}$ for nuclear ring clusters older than 15 Myr is 0.2 magnitude less than for young clusters.

To explore spatial correlation with age, in Figure~\ref{Benedict.fig12} we  map clusters represented by symbols corresponding to the two age groups (age older or younger than 15 Myr). There is a slight tendency for a radial age gradient in that that the few clusters outside the ring are all older than 15 Myr. There is no evidence for a smooth azimuthal gradient. However, most of the clusters with 40\arcdeg $<$ PA $<$ 150\arcdeg~and 180\arcdeg $<$ PA $<$ 260\arcdeg~are young. At all other PA there is a mix of young and old clusters. We note that the dust lanes enter the nuclear ring at approximately the above ranges of PA. The apparent lack of older clusters in these regions could be due to dust extinction lowering their brightnesses below our photometric cut-off (Section~\ref{CUT}).

\subsection{ The V-band Cluster Luminosity Function } \label{VCLF}
Our adopted distance, 13.1 Mpc, and estimated error imply a distance modulus of m-M = 30.59$\pm0.15$.
We correct each cluster V for the derived $A_V$ intrinsic to NGC 4314 and obtain the absolute magnitudes listed in Table 4. These range $-8.0 < M_V <-11.9$, values similar to those found for clusters in NGC 1326 (Buta et al. 2000). 
The most luminous star clusters in NGC 4314 have
corrected absolute magnitudes of $-$11.9, about 2 magnitudes more
luminous than the most luminous star clusters
in the Milky Way (Harris 1991).
These clusters are slightly more luminous
than R136, the cluster at the core of
the giant H~II region 30 Doradus in the Large Magellanic Cloud
(M$_V$ = $-$11.3; O'Connell et al. 1994).
NGC 4314 has no star clusters as luminous as
those found in the interacting galaxies/mergers NGC 1275, NGC 7252,
NGC 4038/9, and NGC 3921 
(Holtzmann et al. 1992;
Whitmore et al. 1993; 
Whitmore $\&$ Schweizer 1995; Schweizer et al. 1996), 
the irregular galaxies NGC 1140, NGC 1569, NGC 1705, 
and M82 (Hunter et al. 1994;
O'Connell et al. 1994, 1995),
or the circumnuclear rings in 
the barred galaxies
NGC 1097, NGC 6951, and ESO 565-11
(Barth et al. 1995; Buta et al. 1999). However, some of the NGC 4314 clusters do fall in the range of the most luminous clusters identified in IC 2163 (Elmegreen et al. 2001).

It is of interest to compare the V band luminosity function (LF)
of the star clusters in NGC 4314 with the luminosity
functions of star clusters in other galaxies studied by {\it HST}.
To accomplish this, we require a more complete V-selected sample
of star clusters than provided by our initial photometric precision culling (Section~\ref{CUT}).  We therefore re-ran the IRAF routine
{\tt daofind} on the V-band image and obtained 154 clusters
with a S/N $\ge$ 10.
We corrected these magnitudes for Galactic and internal
extinction as discussed earlier.
In Figure 13, we plot the distribution of
absolute V magnitudes for these star clusters,
with and without correction for internal extinction.

To investigate the completeness level in our sample, we 
added artificial clusters to the image with a range of apparent magnitudes.
We then ran {\tt daofind} on the resultant image,
and tabulated the percentage of artificial star clusters
at each magnitude level that were successfully recovered.  
These results indicate
that this V-band selected cluster sample is complete to V $\sim$ 22.75, or
an uncorrected M$_V$ of -8.1.
This is just slightly below 
the turn-over in the NGC 4314 LF 
(see Figure \ref{Benedict.fig13}).
Relative to V band photometry only, we are probing 2 $-$ 4 magnitudes fainter 
in the LF
than earlier
studies 
of star clusters
in
NGC 1275 (Holtzmann et al. 1992), NGC 4038/9 (Whitmore $\&$ Schweizer
1995), and
NGC 7252 (Whitmore et al. 1993),
which were based
on images from the now defunct Wide Field Planetary Camera (WFPC).
We are 
0.5 $-$ 1 magnitude more sensitive in the LF 
than the WFPC2 studies of NGC 3921, ESO 565-11, 
NGC 1326, and NGC 7252 (Schweizer et al. 1996; Buta et al.
1999, 2000; Miller et al. 1997), and 3 magnitudes more
sensitive than the WFPC2 NGC 1097 study of Barth et al. (1995). Whitmore et al. (1999) with long exposure WFPC-2 data probe well into the absolute magnitude domain of supergiant stars. They attain an uncorrected completeness limit of V = 24 (M$_V$=-7). As they point out, clusters dominate the LF brighter than M$_V$=-9, while supergiants become important at fainter magnitudes.

We fit the bright end (presumed not severely contaminated by stars) of the uncorrected LF (-8.0 $<$ M$_V$ $<$ -11) to
a power law of form N(L) $\propto$ L$^{\alpha}$dL,
and found a best fit slope of $\alpha$ = $-$1.74 $\pm$ 0.05,
consistent with that found for the interacting pair NGC 4038/9 ($\alpha$ = $-$1.78 $\pm$ 0.5; Whitmore and Schweizer 1995), but
not as steep as that found for other galaxies
(NGC 3921, $\alpha$ = $-$2.1 $\pm$ 0.3; Schweizer et al. 1996), including other nuclear ring galaxies
ESO 565-11 ($\alpha$ = $-$2.18 $\pm$ 0.06; Buta et al. 1999), NGC 1326 ($\alpha$ = $-$2.10 $\pm$ 0.04; Buta et al 2000), and NGC 1512 ($\alpha$ = $-$2; Maoz et al 2001). We note that the LF turns over near M$_V$=-9,
and that the slope at the brightest end from -9.5 to -11
would be steeper, more consistent with other galaxies.

We reiterate that the cluster sample discussed in this section is NOT the cluster sample used in the age estimates above and the following mass estimates. The 76 NRC selected for the photometric study are a sample complete only to M$_V$=-10. 

\subsection{Estimating NRC Masses} \label{MASS}
\cite{Elm99} derive cluster masses by comparing blue absolute magnitudes, M$_B$, and the B-I color index with predictions from cluster models. We have estimated M$_V$ corrected for A$_V$ (Table 4). From M$_V$ and the extinction corrected B-V we obtain M$_B$. We plot in Figure \ref{Benedict.fig14} our cluster M$_B$ vs. B-I, along with values for synthetic clusters of $10^4$ and $10^5$ ${\cal M}_{\sun}$ from SB99. To obtain the mass in each cluster we interpolate between the two synthetic cluster (and the missing $10^3 {\cal M}_{\sun}$) lines. Cluster masses are provided in Table 4 and have a range 
$0.2\times10^4~{\cal M}_{\sun} < {\cal M}_{NRC} < 4.6\times 10^4 ~{\cal M}_{\sun}$ with an average
$<{\cal M}_{NRC}> = 1.6\times 10^4 ~{\cal M}_{\sun}$. Even though many of these clusters are brighter than any Milky Way cluster (Section~\ref{VCLF}), all are less massive than any Milky Way globular cluster. Hence, these are newly formed open clusters, not proto-globular clusters, more luminous because they are young. The total mass of all measured clusters is $11.8\times10^5 {\cal M}_{\sun}$. NGC 1326 (Buta et al. 2000) contains some clusters with masses as high as  $5\times 10^5 ~{\cal M}_{\sun}$. From Maoz \etal (2001) the average mass for clusters in NGC 1512 (within 4 magnitudes of the brightest cluster) is $<{\cal M}> = 0.7\times 10^4 ~{\cal M}_{\sun}$.

To estimate the total mass in all clusters in the nuclear ring of NGC 4314, we assume that 50\% of the 110 objects removed due to large photometric errors produced little flux because they were faint. We assign them the smallest measured mass, $0.2\times10^4 {\cal M}_{\sun}$. We assume the other 55 objects were average in size ($1.6\times 10^4 ~{\cal M}_{\sun}$), but so affected by interstellar absorption that they appeared faint. This yields a total stellar mass in recently formed clusters of ${\cal M}_* = 2.1\times 10^6 ~{\cal M}_{\sun}$. Because some of the clusters are faint due to age, this is a lower limit for the total mass.
 
As discussed in Section \ref{NRCAP} we have not spatially resolved these clusters, and therefore cannot derive stellar densities. There is a critical density below which a cluster is subject to tidal disruption. Clusters below this critical density have relatively short lifetimes (\cite{Bar95}). We can, by comparing with the Barth et al. (1995) results for NGC 1097, estimate the critical density of a cluster in the nuclear ring. From the P3 CO rotation curve we derive the Oort constant A = 220 km s$^{-1}$ kpc$^{-1}$ and $\omega = 0.077$ Myr$^{-1}$ at r = 5 arcsec. Our derived critical density is $18.5 {\cal M}_{\sun}$ pc$^{-3}$, nearly identical to that for NGC 1097,  $20 {\cal M}_{\sun}$ pc$^{-3}$. Presuming cluster sizes similar to NGC 1097 ($\sim3$pc) and an average cluster mass, $<{\cal M}_{NRC}> = 1.6\times 10^4 ~{\cal M}_{\sun}$, the clusters in the NGC 4314 nuclear ring have stellar densities $\sim140 {\cal M}_{\sun}$ pc$^{-3}$, well above the threshold of stability.

\subsection{Nuclear Ring \Ha ~Morphology}
Figure~\ref{Benedict.fig15} shows the distribution of \Ha ~intensity throughout the nuclear ring. This image has been smoothed with a 
Gaussian ($\sigma = 1$ pixel). Comparing with Figure~\ref{Benedict.fig5}, where clusters and dust are most easily seen, there appear to be some instances where the peak \Ha ~emission and the clusters are spatially coincident (e.g., NRC 47) and many cases where they are anti-coincident. Examples of the latter are the groups of clusters NRC 10, 11, 14, and 18; and
NRC 46, NRC 8, and NRC 34, identified in the finder chart, Figure~\ref{Benedict.fig2}. These areas may be \Ha~bubbles possibly formed by past SN events. Examples of these are discussed by \cite{Whi99a} in the NGC 4038/39 collision and in the nuclear ring of NGC 1512 by Maoz et al. (2001). 

We detect no organized diffuse \Ha~emission
associated with the blue spiral arms external to the nuclear ring. In that smoothed image our $3\sigma$ flux limit was 2.2$\times$ 10$^{-19}$ erg s$^{-1}$ cm$^{-2}$ pixel$^{-1}$. This corresponds to a cluster aperture photometry level of 1.3$\times$10$^{-17}$ erg s$^{-1}$ cm$^{-2}$ and an \Ha~luminosity in our 0\farcs4 diameter aperture of $\le3\times10^{35}$ erg s$^{-1}$. Smoothing improved our detectability of diffuse emission by a factor of $\sim7$.

\subsection{Star Formation Rates in the NRC}

We can estimate the star formation rate (SFR) in the nuclear ring from the estimated total nuclear ring cluster mass (Section \ref{MASS}) and the perceived duration of this star formation event (Figure~\ref{Benedict.fig10}). We find that NGC 4314 has formed 2.1 $\times$ 10$^6{\cal M}_{\sun}$ of stars over $\sim2\times10^7$ yr. This yields a SFR $\sim 0.11{\cal M}_{\sun}$ yr$^{-1}$, nearly an order of magnitude less than that of NGC 1326 (\cite{But00}). We would like to compare the rate of star formation associated with clusters to a global rate from a total $H\alpha$ flux. As discussed in Section \ref{PHOT}, our total $H\alpha$ flux was measured to be $\sim4 \times$ 10$^{-13}$ 
erg s$^{-1}$ cm$^{-2}$, uncertain due to [N{\sc ii}] contamination. Correcting this flux for an average \Ha~absorption derived from from Table 4 (see Section \ref{HaNRC}),  we obtain log L$_{H\alpha}$ = 40.22 and a global SFR $= 0.13{\cal M}_{\sun}$ yr$^{-1}$. These rates suggest that 70-100\% of the star formation in NGC 4314 occurs in clusters. Zepf et al. (1999)
and Meurer et al. (1995) find values of $\sim20$\% for the fraction of star formation in clusters for starburst/merger galaxies.

We note that the total H$_2$ mass inferred for NGC 4314 from the P3 CO study is
$2.1 \times 10^8 {\cal M}_{\sun}$, about half that of NGC 1326 (\cite{But00}). Hence, there is not a linear relationship between the amounts
of gas and star formation, and the much lower SFR in NGC 4314 cannot be caused
simply by a smaller supply of raw material.

\subsection{Notes on Individual NRC} \label{NOIC}
\begin{figure}
\epsscale{1.0}
\plotone{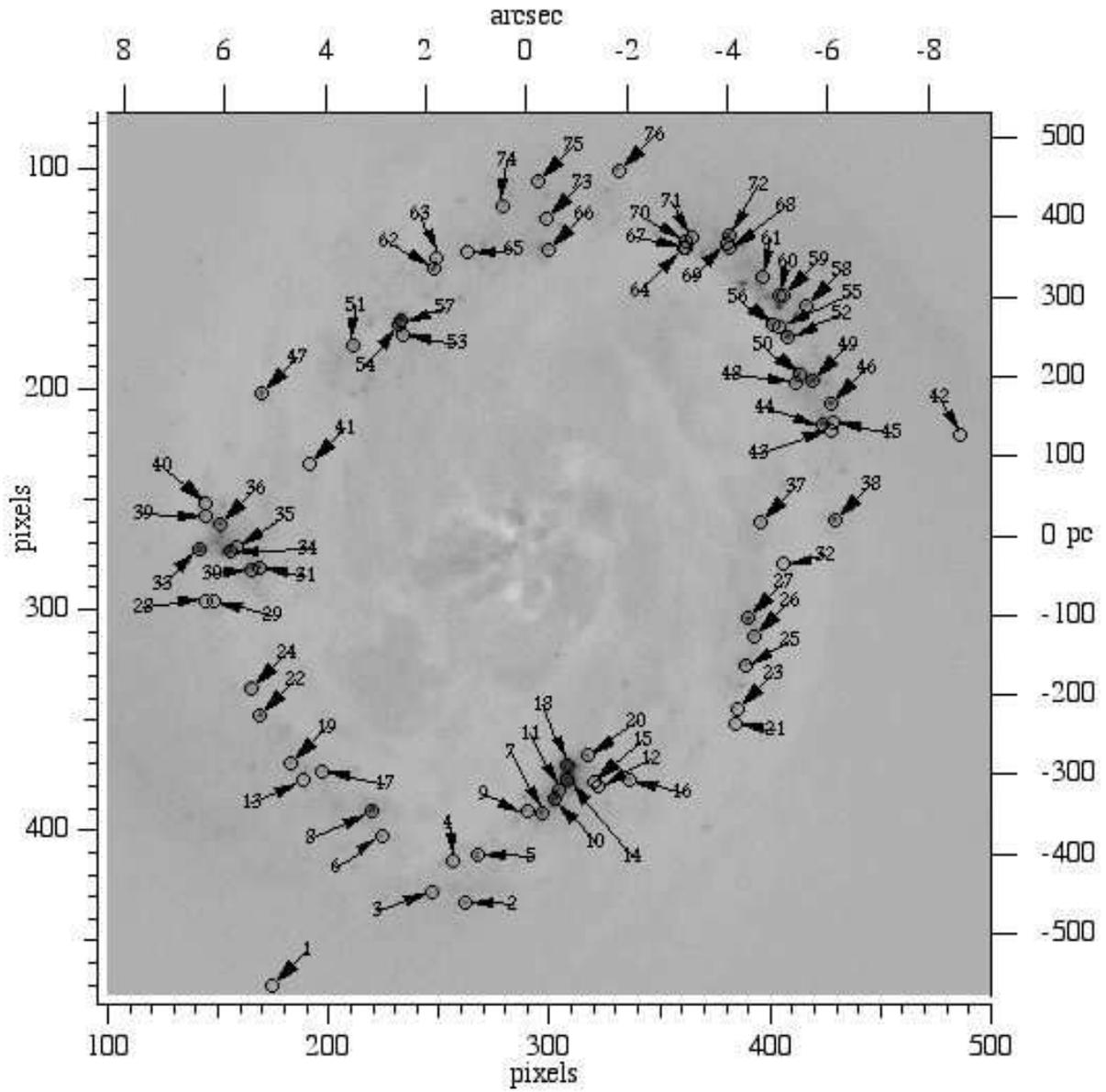}
\caption{Finding chart for the selected clusters (Section~\ref{CUT}) associated with the nuclear ring (NRC) and listed in Table 2.  North is at top, east to the left. We number the clusters in declination order.} 
\label{Benedict.fig2}
\end{figure}

\begin{figure}
\epsscale{0.8}
\plotone{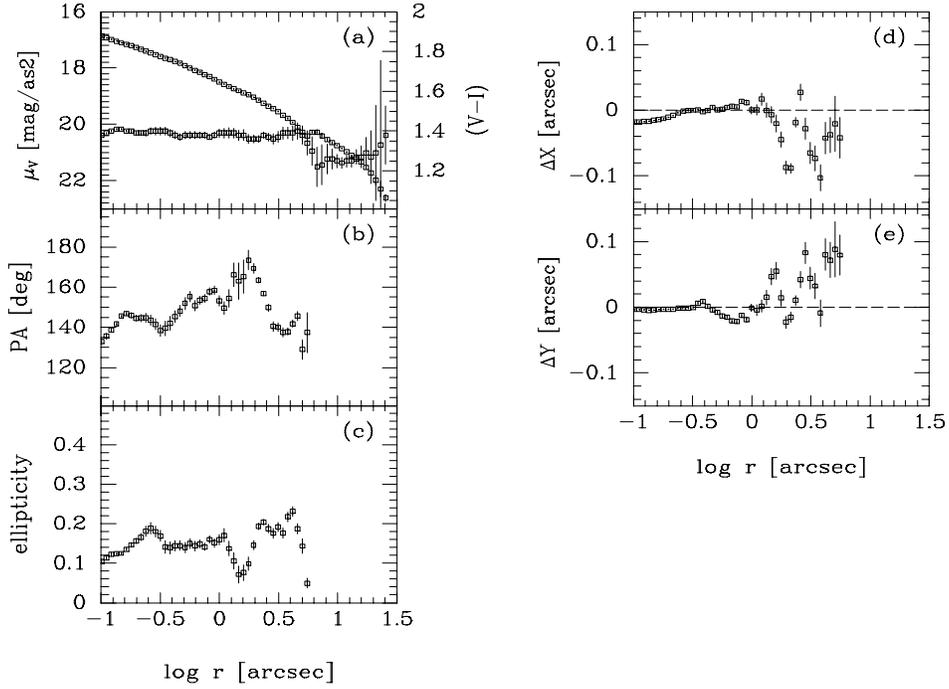}
\caption{Results from fitting ellipses to the I-band isophotes. All parameters
are shown as a function of the equivalent radius of the ellipse ($r=\sqrt{a \times b}$).
(a) --  Surface brightness in V and local color (V-I), (b) --  Position angle of the best fitting ellipse, North through East, (c) --  Ellipticity of the best fitting ellipse, (d) --  Change in ellipse center in X, (e) --  Change in ellipse center in Y. The error bars are in all cases one standard deviation.} 
\label{Benedict.fig3}
\end{figure}

\begin{figure}
\epsscale{1.0}
\epsscale{0.8}
\plotone{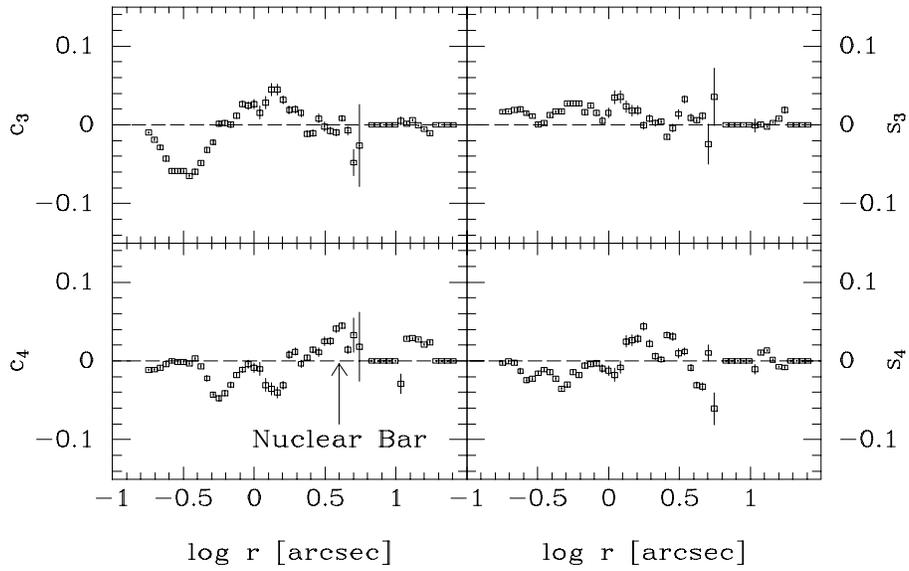}
\caption{Results from fitting ellipses to the I-band isophotes. This figure shows
the deviations of the isophotes from the best fitting ellipses. The
deviations are indicated by the Fourier coefficients $c_3$, $s_3$, $c_4$ and
$s_4$. The error bars are in all cases one standard deviation.
The approximate position of the nuclear bar is indicated on the $c_4$ panel.} 
\label{Benedict.fig4}
\end{figure}

Some of the nuclear ring clusters have interesting extremes. 
All can be located using the finding chart, Figure~\ref{Benedict.fig2}. The discussed values are from Tables 2, 4, and 5.

The most reddened cluster (A$_V$ = 2.26 $\pm$ 0.18) is NRC 24. Identifying it on Figure~\ref{Benedict.fig2}, we see in Figure~\ref{Benedict.fig5} that it is one of the few clusters lying on a dust lane. Most are off to one side or the other of the most prominent dusty regions.
In the Q vs \bv diagrams (Figure~\ref{Benedict.fig7}) and the Q-Q$_3$ diagram (Figure~\ref{Benedict.fig10}) one cluster, NRC 53, is clearly younger than any other. It has the most negative Q value and sits very near to the 1 Myr age marker. 
NRC 57 is the cluster with the largest \Ha ~luminosity (L(\Ha) = 38.46 ergs s$^{-1}$), the brightest absolute V magnitude ($M_V$=-11.88$\pm$0.06), and hence, one of the largest cluster masses (3.8$\times10^4~{\cal M}_{\sun}$). The most massive cluster is NRC 11 with ${\cal M}=4.6\times10^4~{\cal M}_{\sun}$.

NRC 53 and NRC 57 are within 25 pc of each other. NRC 53 suffers nearly 1.5 magnitudes more extinction than NRC 57. Correcting for this extinction, they have similar masses. The younger, NRC 53, has less than half the extinction-independent \Ha~equivalent width of the older, NRC 57.
It is a theoretical possibility that observations at this epoch are temporally resolving the growth of an \ion H 2 region,
catching NRC 53 before the Str\o mgren sphere expands to its maximum surface area and brightness. However, it is more likely that
the younger cluster has more of its ionizing radiation absorbed by dust. The older cluster has probably emerged more from the dust
associated with its birth cloud.

The oldest clusters on Figure~\ref{Benedict.fig10} are NRC 1, NRC 76, NRC 37, and NRC 42, with ages $\sim$ 50 Myr.  The average \Ha ~luminosity for the two oldest NRC with detectable signal is log L(\Ha) = 36.61  ergs s$^{-1}$. Two of these oldest NRC are located well outside the nuclear ring: NRC 1 and NRC 42. NRC 1 is $\sim$3\arcsec~exterior to and SE of the nuclear ring, further than any other cluster from the nuclear ring of active star formation. NRC 42 is within the NW blue
arm. It was the only discrete clump found in either blue arm by our cluster identification process. It is also the faintest cluster to make our photometric precision cut, with 
$M_V$ = -8.06$\pm$0.37. For NRC 42 there is very little associated absorption (A$_V$=0.27$\pm$0.31) and a null detection of \Ha ~emission (0.12$\pm0.11 \times10^{-15}$ ergs s$^{-1}$ cm$^{-2}$). Q-Q$_3$ (Figure~\ref{Benedict.fig10}) indicates an age $\sim$40 Myr with a substantial uncertainty due to the photometric errors and
the clumping of ages 20 through 50 Myr on the Q-Q$_3$ plane. The age of NRC 42 is actually somewhat better constrained in the color-color diagrams shown in Figure~\ref{Benedict.fig9}. NRC 42 lies very near the 50 Myr marker in both diagrams. It is the only resolved cluster in our sample, with a FWHM = 4 pixels = 11.2pc.

\section{Estimating Ages for Stars in the Blue Arms}
\begin{figure}
\epsscale{1.0}
\plotone{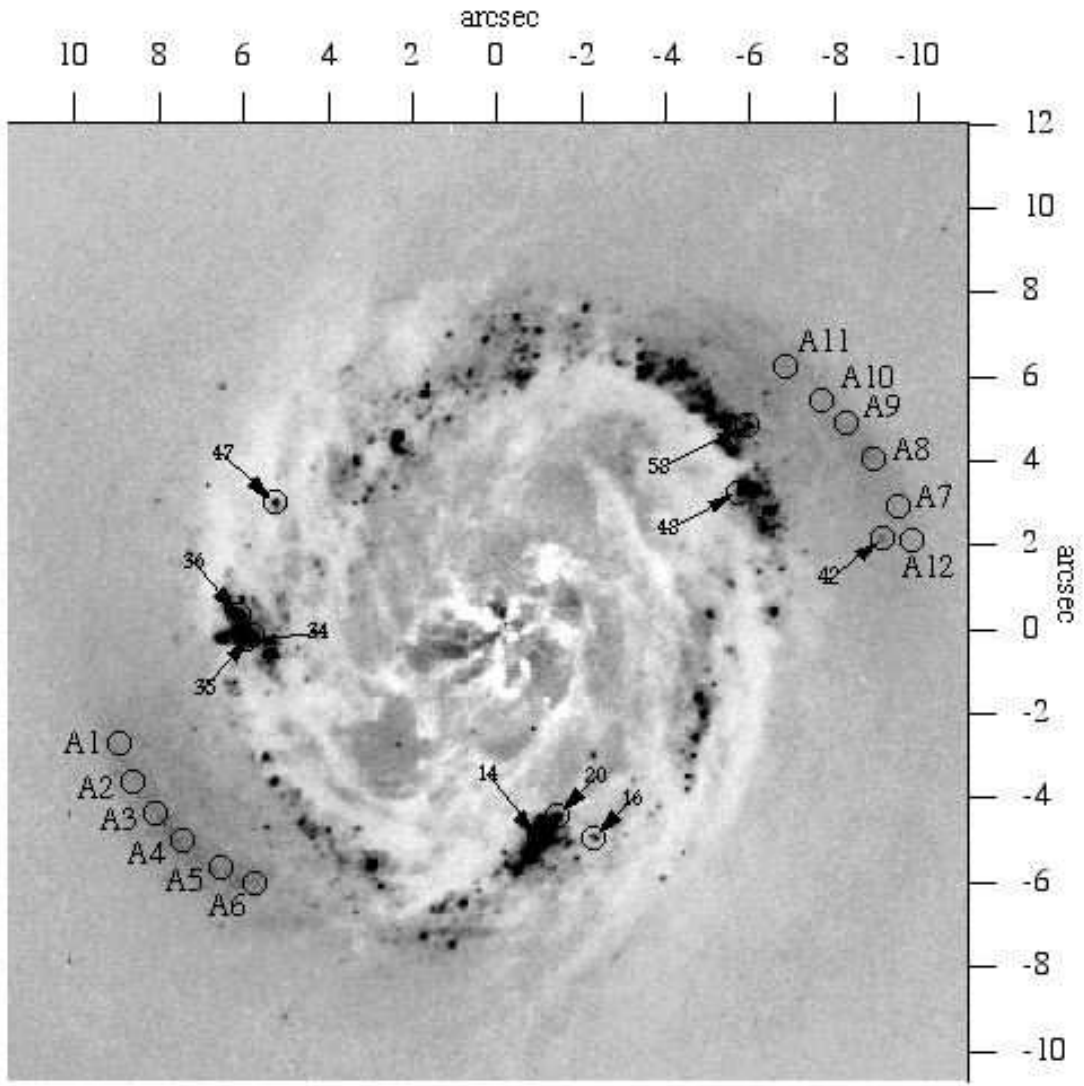}
\caption{Deviations from the nested ellipse model. This image is the sum of all residual maps; U, B, V, and I. Regions with intensity in excess of the model are encoded dark (clusters and outer arms), regions below the model level are white (dust). North is at top, east to the left. A1 - A12 are locations within the blue arms for surface photometry. A few clusters within the ring of active star formation are also indicated, as is NRC 42, the only cluster identified in the blue arm region.} 
\label{Benedict.fig5}
\end{figure}

\begin{figure}
\epsscale{0.6}
\plotone{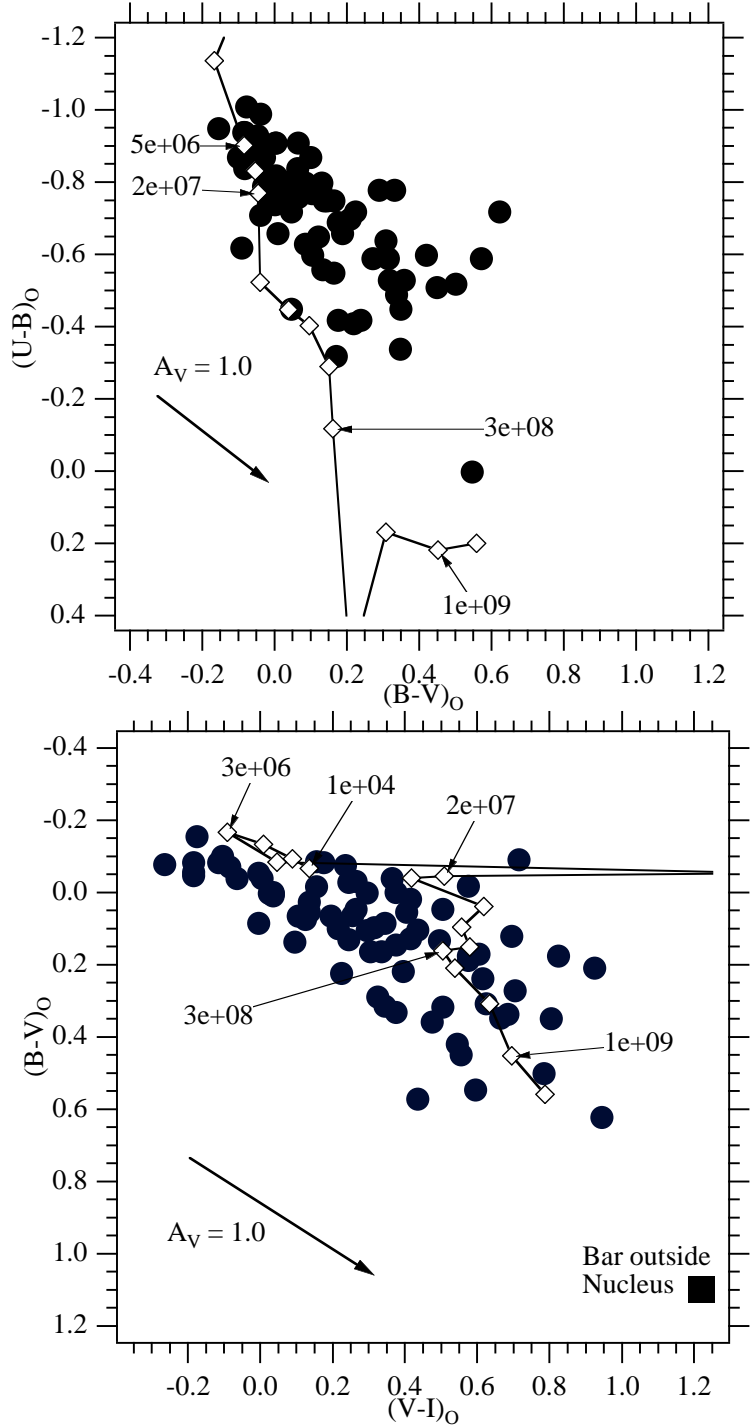}
\caption{Color-color diagrams (top: (\ub)$_O$ ~vs. (\bv)$_O$; bottom: (\bv)$_O$~vs. (\vi)$_O$) for photometry of the NRC in Table 2. The symbol size is that of the median 1$\sigma$ photometric error. Also plotted are a reddening vector for A$_V$ = 1.0 and a cluster evolution track ($\Diamond$) from SB99 with a few associated ages in years. The square indicates the (\bv)$_O$ and (\vi)$_O$ ~colors for the primary stellar bar exterior to the nuclear ring region and for the nuclear bar at location x=-1\farcs5, y=+3\farcs0 in Figure~\ref{Benedict.fig5}.} 
\label{Benedict.fig6}
\end{figure}

\begin{figure}
\epsscale{1.0}
\epsscale{0.5}
\plotone{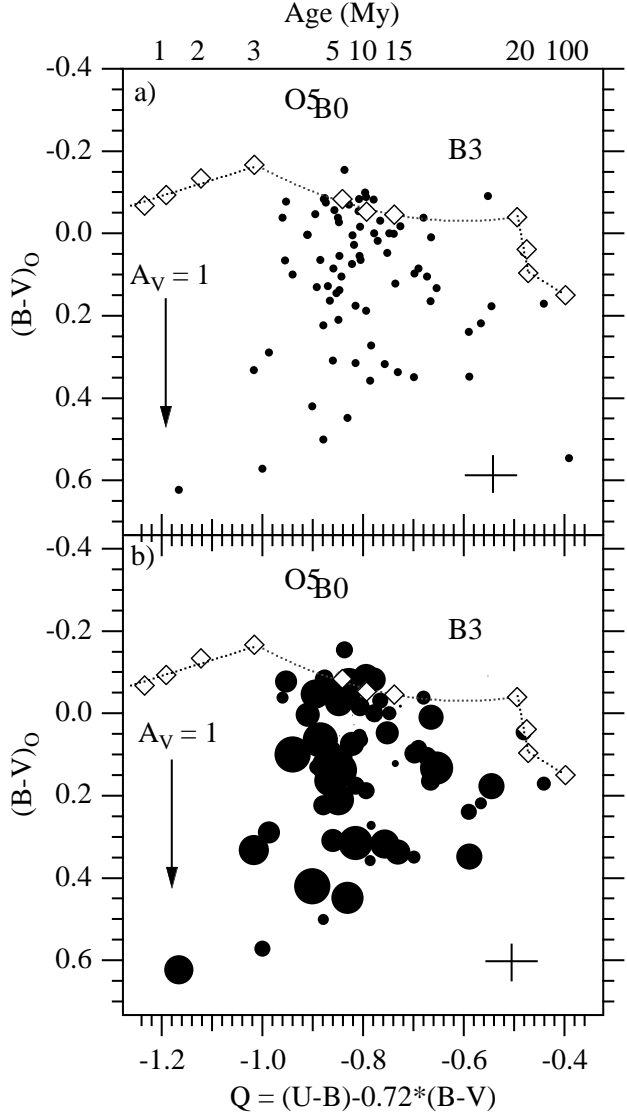}
\caption{a) (\bv)$_O$  plotted against the reddening-free parameter, Q. Also plotted are positions of main sequence stars O5 through B3, a reddening vector for A$_V$ = 1.0, and a cluster evolution track ($\Diamond$) from SB99 with associated ages in years along the top axis. Left to right, specific cluster ages are 0.1, 1, 2, 3, 5, 10, 15, 20, 30, 50, and 100 Myr. The cross provides median 1$\sigma$ photometric errors. b) (\bv)$_O$ vs. Q, where now the symbol size is proportional to an \Ha~equivalent width index, $EW_{H\alpha}$. The largest symbols have an $EW_{H\alpha}$ $\sim2.5$ magnitude larger than the smallest.} 
\label{Benedict.fig7}
\end{figure}

\begin{figure}
\epsscale{1.0}
\plotone{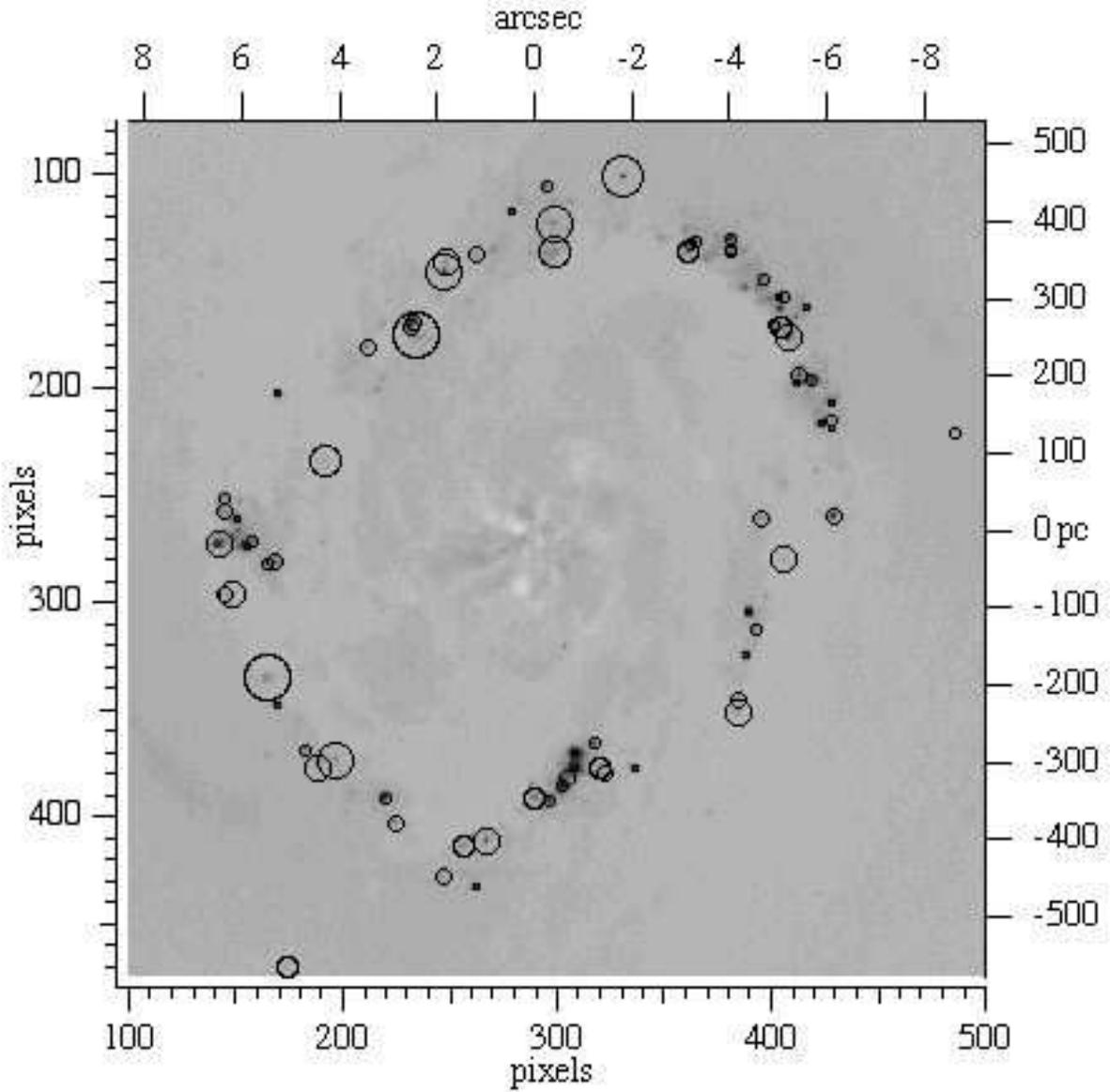}
\caption{NRC $A_V$ on a shallower stretch of the Figure 5 nested ellipse deviation map. Symbol size is proportional to $A_V$. The largest extinctions ($A_V$=2.2) are found to the NE, the far side of the nuclear ring.}
\label{Benedict.fig8}
\end{figure}

\begin{figure}
\epsscale{0.6}
\plotone{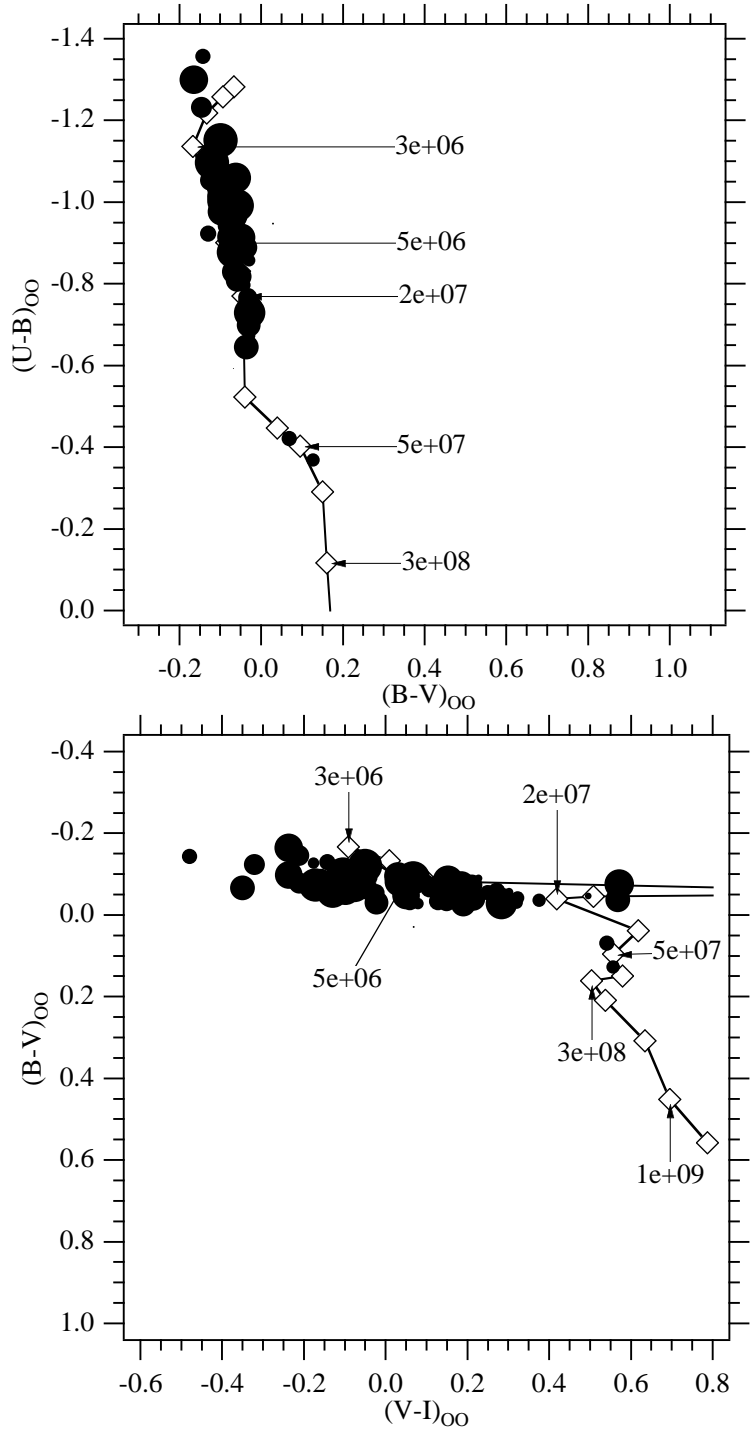}
\caption{Color-color diagrams (top: (\ub)$_{OO}$ ~vs. (\bv)$_{OO}$; bottom: (\bv)$_{OO}$~vs. (\vi)$_{OO}$) for the clusters comprising the nuclear ring, now corrected for extinction within NGC 4314. Compare with Figure~\ref{Benedict.fig6}, where colors are uncorrected for internal extinction. The symbol size is proportional to an extinction-independent \Ha~equivalent width index, $EW_{H\alpha}$, larger circles for larger EW. Also plotted is a cluster evolution track ($\Diamond$) from SB99 with a few associated ages in years. The extinction corrected cluster loci are inconsistent with a single, brief epoch of star formation.} 
\label{Benedict.fig9}
\end{figure}

\begin{figure}
\epsscale{0.7}
\plotone{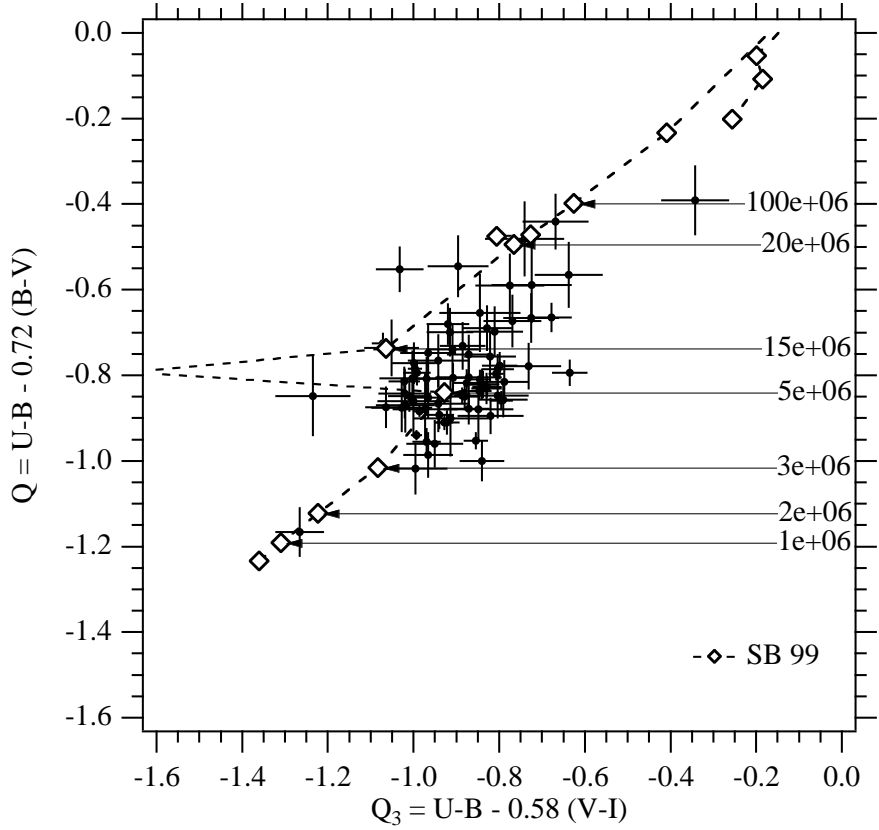}
\caption{Q vs. Q$_3$.  Errors are $1\sigma$. Also plotted are the corresponding values for model clusters from SB99 ($\Diamond$) with ages in years along the right axis. Nuclear ring clusters clump at 5 Myr. The youngest cluster is NRC 53, the oldest NRC 76. NRC 42 (in the blue arm to the West) is just above the 20 Myr model cluster symbol.
} 
\label{Benedict.fig10}
\end{figure}

\begin{figure}
\epsscale{0.7}
\plotone{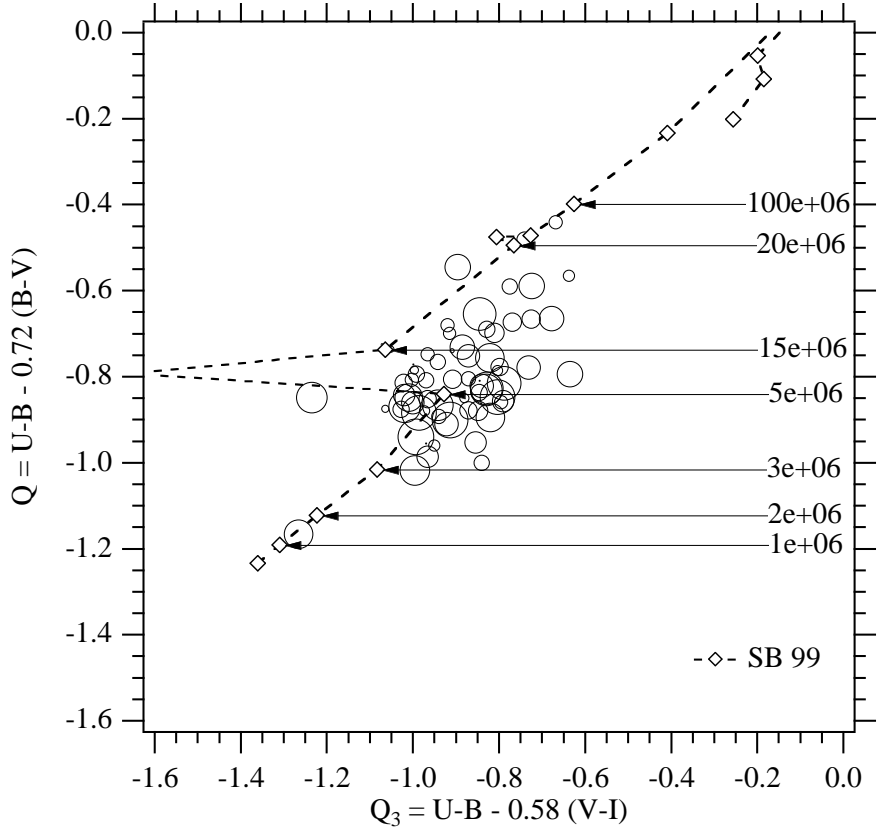}
\caption{Q vs. Q$_3$. Also plotted are the corresponding values for SB99 ($\Diamond$) with ages in years along the right axis. The symbol size is proportional to the extinction-independent \Ha~equivalent width index, $EW_{H\alpha}$. 
\Ha~flux and age are expected to correlate inversely. NRC 42 and others have disappeared, having essentially no associated \Ha~flux.} 
\label{Benedict.fig11}
\end{figure}

We return to the region just external to the nuclear ring, the blue arms. These were first discussed in P1 and confirmed in the photometry of Wozniak et al. (1995). We found only one cluster in these arms (NRC 42) and detect no diffuse \Ha~associated with the arms or \Ha~flux from NRC 42. From the age of NRC 42 ($\sim$50 Myr) and
the lack of \Ha~emission associated with this cluster and the arms in general (Figures~\ref{Benedict.fig10} and \ref{Benedict.fig15}) we estimate that the arms are older than
the ring of active star formation.  From P3 we conclude that there is little CO (see Figure~\ref{Benedict.fig16}, combining the CO intensity map from P3 with the nested ellipse model residual map), hence little reddening and little raw material for ongoing star formation. From a CO - A$_V$ relation (P3, figure 11) none of the sample regions A1 - A12 (Figure~\ref{Benedict.fig5}) should suffer more than A$_V$= 0.03 absorption. 

In Figure~\ref{Benedict.fig17} we present color-color diagrams for the selected ring clusters and  blue arm photometry at the Figure~\ref{Benedict.fig5} locations. These colors are obtained by subtracting a modeled, smoothly varying  background (Section~\ref{SPBA}). Presuming no extinction, we can determine the region ages, 100-200~Myr. Unlike the clusters in the active ring of star formation, the light from the blue arms is evidently not dominated by
young, massive blue stars. If this is a single age population with an age of $\sim200$ Myr, those arms are now a mix of evolved red giants (whose progenitors were B8 and earlier) and unevolved stars of spectral type B9 and later.

To provide another estimate of the age of the stars intrinsic to the blue arms we calculate Q and Q$_3$ (equations 2 and 4) for the surface photometry of regions A1 - A12 and plot them in Figure~\ref{Benedict.fig18}, along with the SB99 model clusters of various ages. Again plotted are a few clusters from the ring of active star formation (see the finding chart, Figure 5). Inferred ages for the stars comprising the blue arms (from the surface photometry listed in Table 3) range from 20 Myr for location A6 to $\sim$200 Myr for location A4. There is weak evidence of an age progression along each arm. Comparing Figures~\ref{Benedict.fig18} and \ref{Benedict.fig5} the locations further from the nuclear ring (A1, A2, A4, A7, A8, A9, A12) are likely older than 100~Myr. Locations in the Q - Q$_3$ plane suggest ages less than 100 Myr for regions closer to the nuclear ring (A3, A5, A6, A10). 
NRC 42, well within the western blue arm, very near  A12, has no detectable \Ha ~flux  and an age of ~50 Myr. This one NRC certainly violates the age progression along the blue arm. The photometric quality of these data do support the assertion that in general the arm stars are older than the stars in the ring clusters. They only weakly support an age gradient along the arms.

\section{Discussion}

Compared to NGC 1326 (\cite{But00}) and NGC 1512 (Maoz \etal 2001), the nuclear ring in NGC 4314 has an \Ha~luminosity (hence an inferred SFR) a factor of ten less, but in contrast to these galaxies has many clusters with directly associated \Ha~flux. The cluster formation process in NGC 1326 builds some clusters ten times more massive than the largest in NGC 4314. NGC 1512 produces clusters with a mass distribution similar to NGC 4314. NGC 4314 has half the internal extinction of NGC 1326, but about the same as NGC 1512. Each of these galaxies has produced clusters with a similar absolute magnitude range. Neither NGC 1326 nor NGC 1512 has circumnuclear blue arms, sites of past star formation just exterior to their nuclear rings. All three nuclear rings contain a similar small fraction of clusters older than 100 Myr.

\begin{figure}
\epsscale{1.0}
\plotone{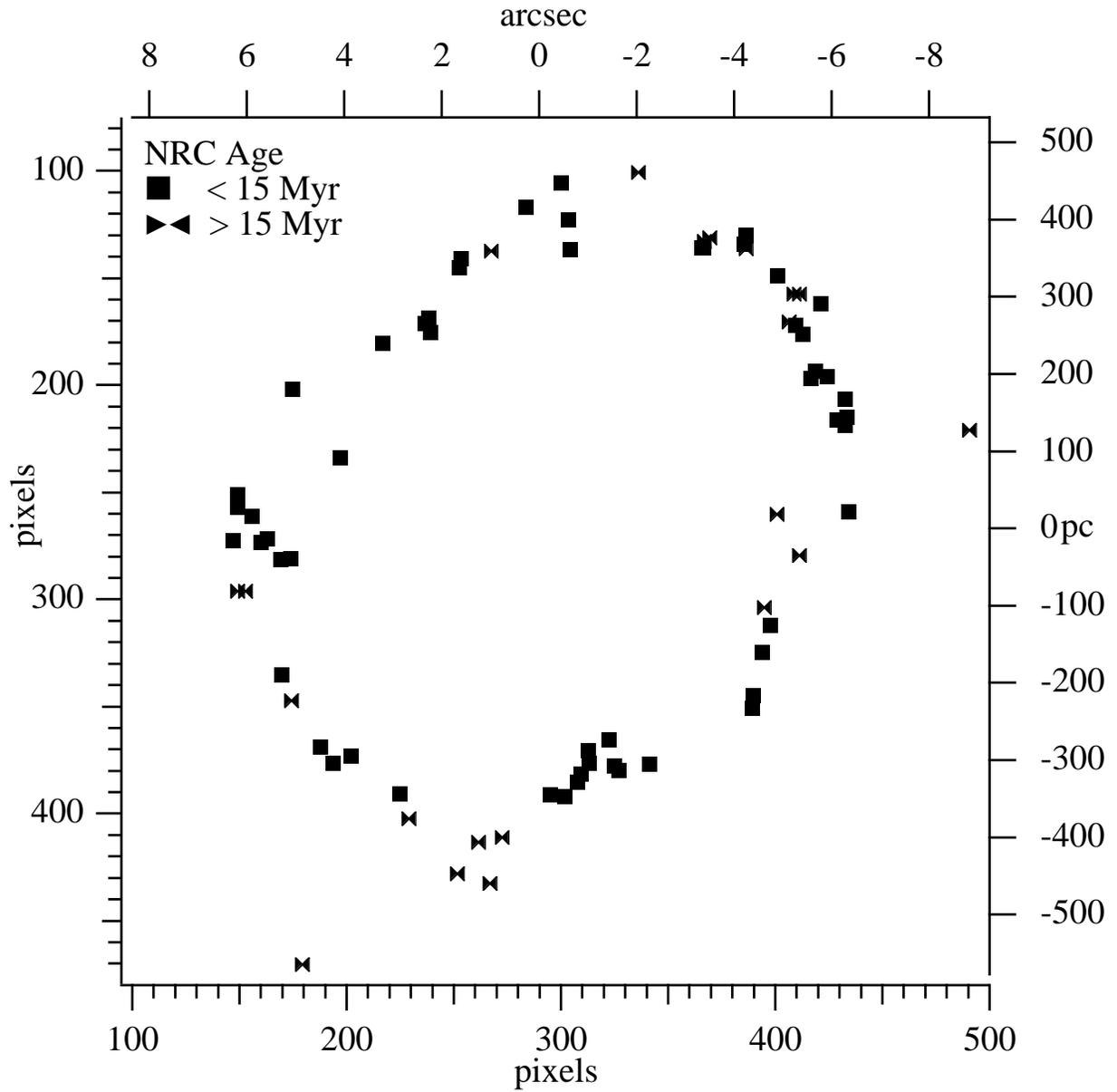}
\caption{Nuclear ring clusters identified by age. Filled boxes denote Age$<$ 15 Myr and bow-ties, Age $>$ 15 Myr. Note the lack of an azimuthal age gradient and weak
evidence for a radial age gradient. } 
\label{Benedict.fig12}
\end{figure}

\begin{figure}
\epsscale{0.7}
\plotone{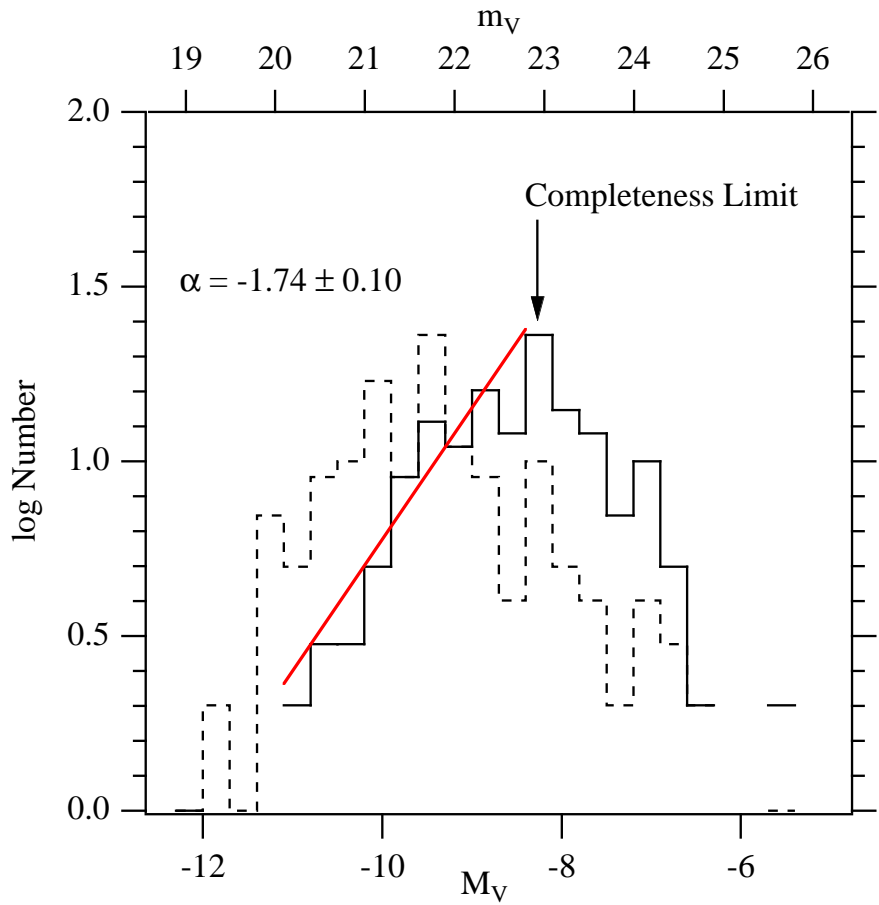}
\caption{The V-band luminosity function.  
The dashed line shows the distribution of absolute V magnitudes
after correction
for internal extinction; the solid line is the distribution of uncorrected
magnitudes.
} 
\label{Benedict.fig13}
\end{figure}

\begin{figure}
\epsscale{1.0}
\plotone{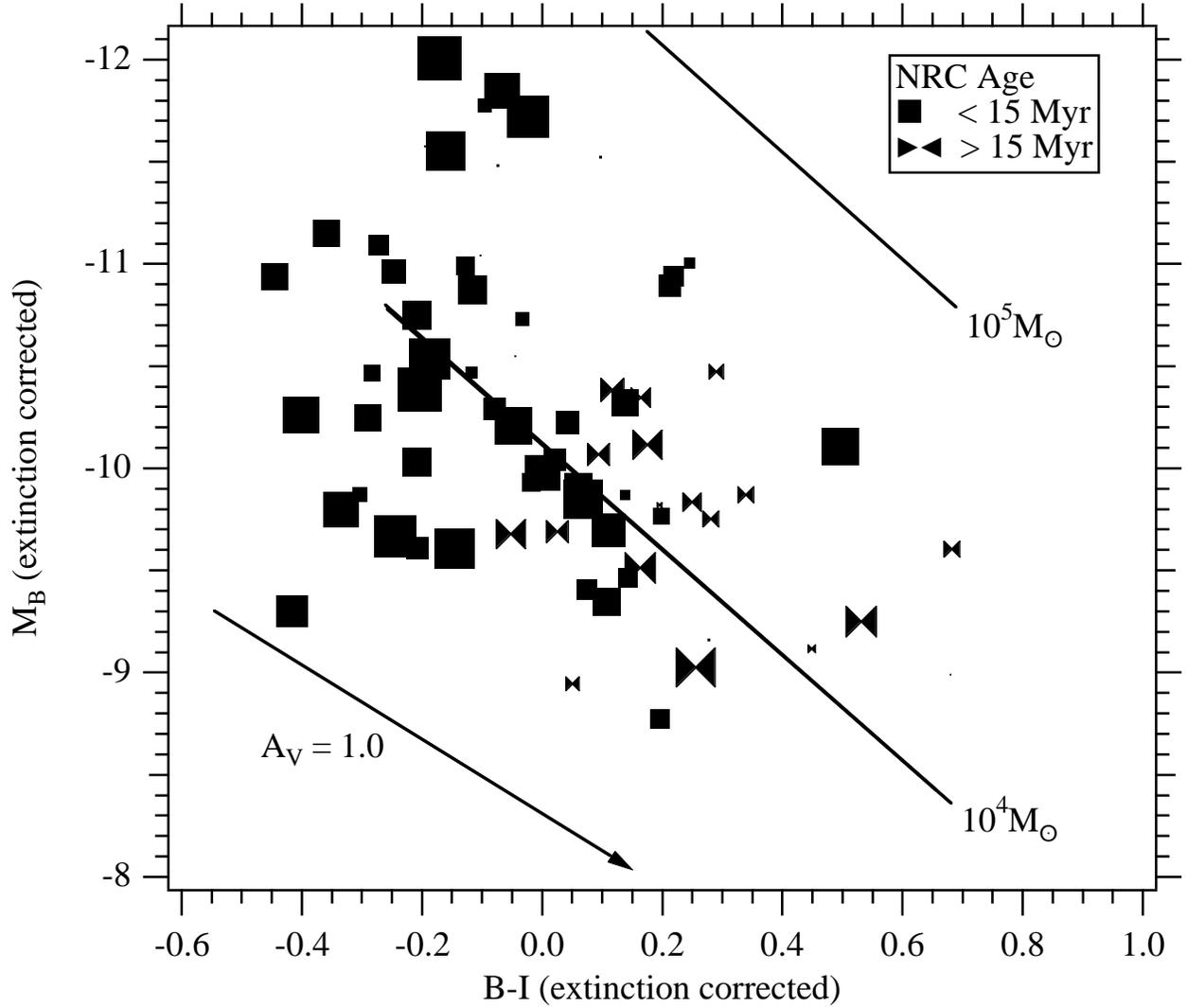}
\caption{Estimating cluster mass from blue absolute magnitude, M$_B$, and B-I. Both have been corrected for internal absorption. Also plotted are SB99 evolutionary trajectories for clusters with masses $10^4$ and $10^5 {\cal M}_{\sun}$. Filled boxes denote Age$<$ 15 Myr and bow-ties, Age $>$ 15 Myr. The symbol size is proportional to the extinction-independent \Ha~equivalent width index, $EW_{H\alpha}$.} 
\label{Benedict.fig14}
\end{figure}

\begin{figure}
\epsscale{1.0}
\plotone{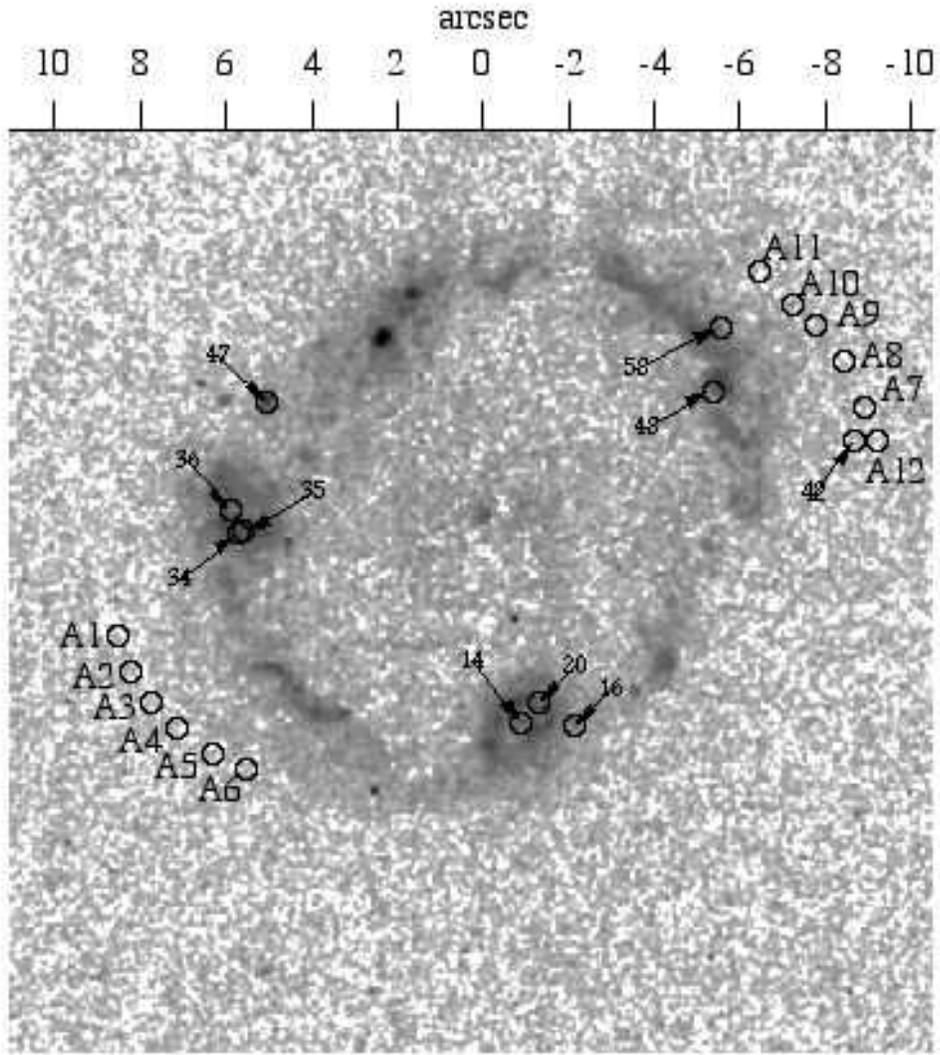}
\caption{Uncalibrated \Ha~+ [N{\sc ii}] map, smoothed with a Gaussian of $\sigma = 1$ pixel. The aperture locations within the blue arms (Table 3) and representative comparison NRC (from Table 2) are marked.
Note possible outflow bubbles in the nuclear ring at the location of NRC 14 and 20. North is at top, east to left.} 
\label{Benedict.fig15}
\end{figure}

\begin{figure}
\epsscale{1.0}
\plotone{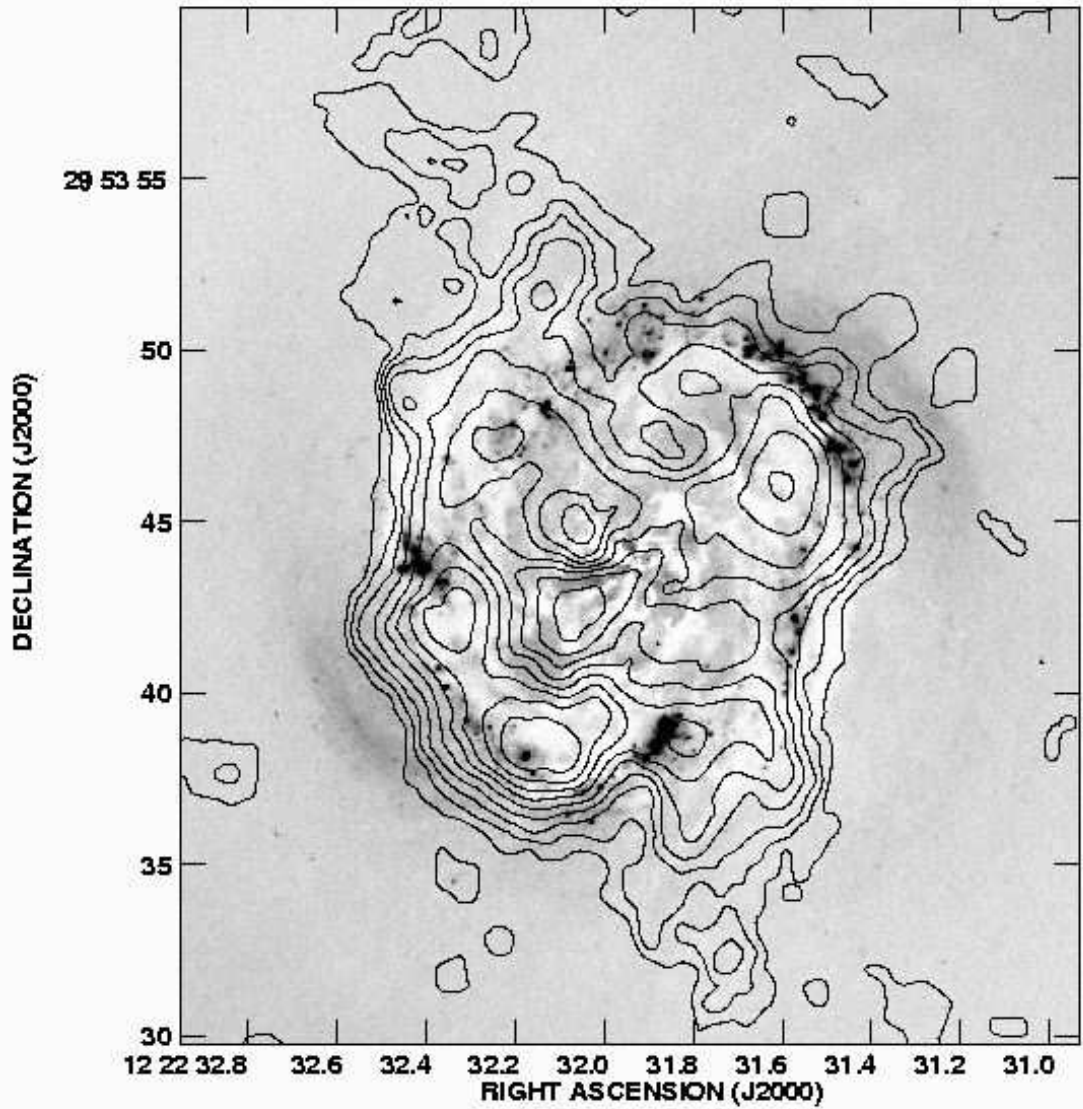}
\caption{CO intensity contours (beam size 2\farcs3$\times$2\farcs2) from P3 plotted on the Figure~\ref{Benedict.fig5} nested ellipse model deviations. The contour levels are 0.1, 0.2, ..., 1.0 times the peak intensity of 1.3 Jy beam$^{-1}$ \kms. From the P3 CO - A$_V$ relation none of the regions A1 - A12 (identified on Figure~\ref{Benedict.fig6}) should suffer more than A$_V$= 0.03 absorption.} 
\label{Benedict.fig16}
\end{figure}

\begin{figure}
\epsscale{0.5}
\plotone{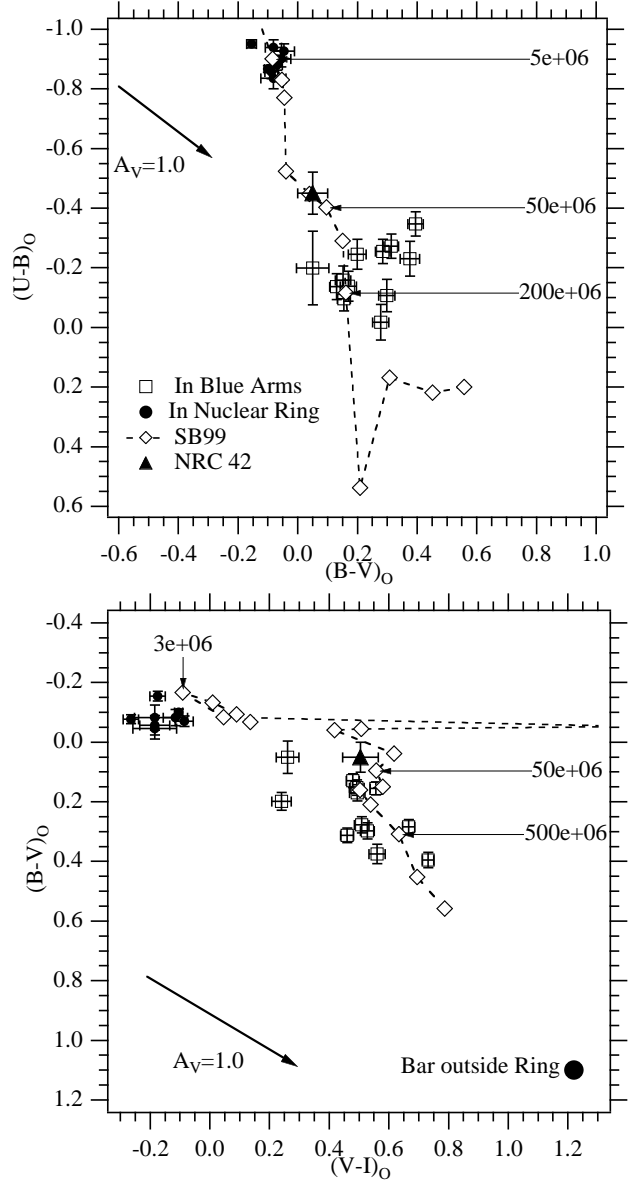}
\caption{ Color-color diagrams (\ub)$_O$ ~vs. (\bv)$_O$~ and (\bv)$_O$ ~vs. (\vi)$_O$ for a few of the selected ring clusters and for blue arm surface photometry regions identified in Figure~\ref{Benedict.fig5}. The blue arm data are surface colors. Also plotted are the SB99 cluster evolution trajectory, the position of NRC 42, and a reddening vector for A$_V$ = 1.0. Errors are $1\sigma$.} 
\label{Benedict.fig17}
\end{figure}

\begin{figure}
\epsscale{1.0}
\plotone{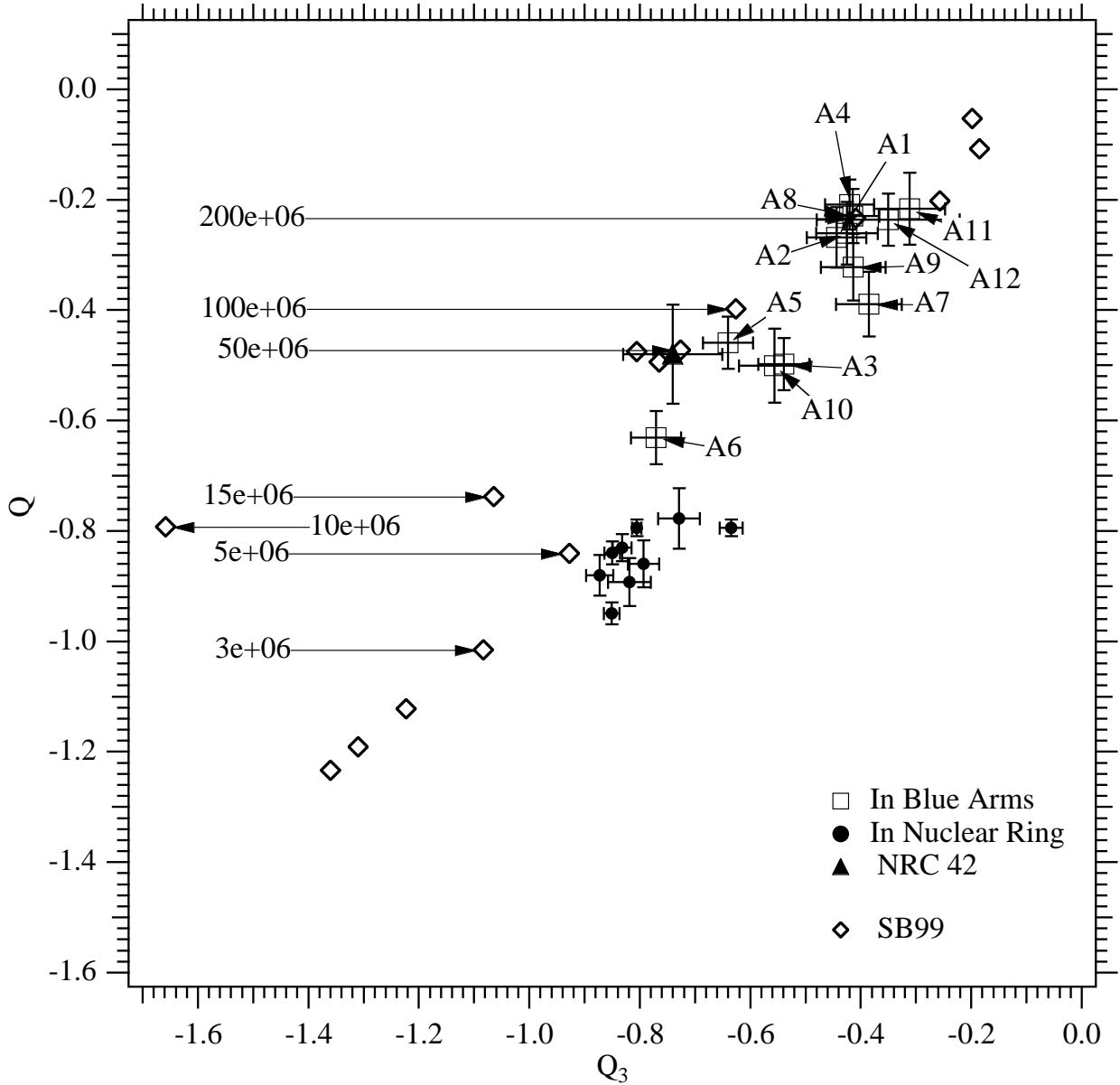}
\caption{Q vs. Q$_3$ for the blue arm regions A1 - A12 (identified on Figure~\ref{Benedict.fig5}) and a few selected clusters associated with the nuclear ring, including the one identified cluster in a blue arm, NRC 42. Errors are $1\sigma$. Diamonds show the corresponding values for SB99 with ages in years along the left axis. These particular NRC clump in age at 5 Myr. Most of the blue arm regions appear to have ages 100-300 Myr. Regions A3, A5, A6, and A10 appear intermediate in age.} 
\label{Benedict.fig18}
\end{figure}

We next discuss the locus of the NRC, the nuclear ring of active, present-day star formation. In P3 we mapped the inflow of gas along the primary stellar bar into the nuclear ring. We find some evidence that the inflow triggers star formation at the locations where gas enters the ring (PA $\sim80\arcdeg$ and $\sim260\arcdeg$, Figure~\ref{Benedict.fig12}). However, the youngest cluster, NRC 53, is, galactic rotation-wise, 'upstream' from the PA$\sim80\arcdeg$ 'impact' site. Star formation does not seem to work its way around the ring in a sequential fashion, like a burning fuse (there is no strong azimuthal age gradient). Neither do we find a significant radial age gradient at any position angle within the nuclear ring. However, the three oldest clusters are at or beyond its outer edge (Figure~\ref{Benedict.fig12}). At the average radius of the NRC distribution (from P1) we identified an IILR  from the CO dynamics discussed in P3. Hence, the star formation mechanism
is consistent with  gravitational instability associated with an IILR (e.g. Elmegreen 1994), which triggers formation simultaneously all around the ring. 

Our sample of clusters in the nuclear ring of NGC 4314 contains no old (age $>$ 100 Myr) clusters (Section~\ref{QQ3HaA}). In NGC 1326 (\cite{But00}) the same age estimation technique showed that 10\% of the nuclear ring clusters were older than 100 Myr. Fitting SED's to models, \cite{Mao01} finds that 4\% of the nuclear ring clusters in NGC 1512 are old. Unfortunately, we cannot make a direct comparison with these results. The NGC 1326 and NGC 1512 cluster samples were defined by fainter M$_V$ limits. Assuming NGC 4314 to have a similar mix of old and new clusters, we should find between 3 and 8 with age $>$ 100 Myr. These may have been missed because they were faint and discarded in our cluster selection process (Section~\ref{CUT}). SB99 estimates that for a given mass, a 100 Myr old cluster would be $\sim$ 2.5 magnitudes fainter than a cluster at age 5 Myr. Our NRC sample contains photometry for only 12 clusters that are 2.5 magnitudes fainter than our brightest cluster. Age alone could reduce nuclear ring clusters in brightness enough to be missed entirely in our sample. Maoz \etal (2001) argue that the ring in NGC 1512 could contain many more old clusters that were too faint to allow detection. The existence of such an older population would argue for continuous (as opposed to episodic) star formation in the nuclear ring. 

Given the importance of the presence or lack of old clusters in the nuclear ring, we inspected a color-color diagram, (\ub)$_O$ ~vs. (\bv)$_O$, for the full initial sample of 186 clusters. It suggests that a small fraction of the NRC in NGC 4314 is older than 100 Myr. Because these are the faintest clusters in our sample with the least precise photometry, it is impossible to derive a precise fraction. We estimate that the fraction of old clusters in the nuclear ring of NGC 4314 is closer to the 4\% found by \cite{Mao01} for NGC 1512, than the 10\% found by Buta et al (2000) for NGC 1326.

Combes et al. (1992) assert that the CO ring (Figure~\ref{Benedict.fig14}) is shrinking due to dynamical friction. Because the CO traces the raw material required for star formation, this shrinking could produce an age gradient across the nuclear ring of active star formation. The photometric quality of our data do not permit a precise determination of an age gradient across the $\sim$1\arcsec~ diameter ring annulus. This would provide a rate for the shrinking. However, if we interpret the scatter in age in Figure~\ref{Benedict.fig10} as an upper limit
to a radial age gradient (as the ring shrinks by~$\sim$1\arcsec), we infer an upper limit for the shrink rate of 64pc/30Myr = 2.1pc/Myr. 

\begin{figure}
\epsscale{1.0}
\plotone{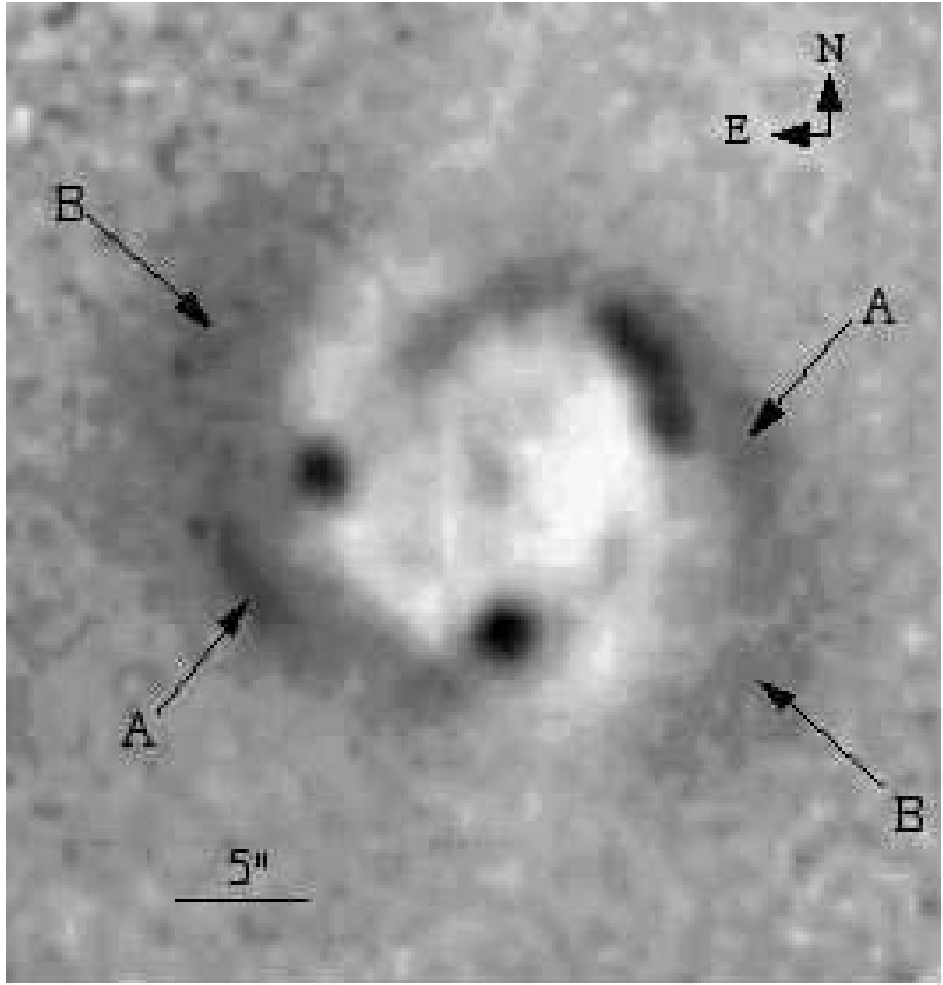}
\caption{The ground-based \vi~map from P1 encoded with darker as bluer. The blue arms discussed in Section 7 are marked `A'.
The remaining parts of the elliptical distribution previously identified with an OILR are marked `B'. The `B' sections are slightly less blue and slightly more dispersed than the `A' sections.} 
\label{Benedict.fig19}
\end{figure}

The blue arms exterior to the nuclear ring of active star formation are  morphological features found neither in NGC 1326 (\cite{But00}) nor NGC 1512 (Maoz \etal 2001). Our {\it HST} blue arm data cannot be discussed without reference to the entire structure within which they lie. One can see in the surface color maps in P1 (the P1 \vi~map is reproduced as Figure~\ref{Benedict.fig19}) and the B-I map in Wozniak et al. (1995) extensions (further to the NW for the SE arm, further to the SE for the NW arm) to the blue arms visible in Figures~\ref{Benedict.fig1} and~\ref{Benedict.fig5}. The blue arms are marked `A' and the extensions marked `B' in Figure~\ref{Benedict.fig19}. The `B' regions are too faint to detect in the {\it HST} data. In Figure~\ref{Benedict.fig19} they are slightly less blue and somewhat broader than the `A' sections. Based partly on the CO rotation curve, we identified in P3 the 
likely location of the OILR near the outer edge of the blue region
containing features A and B.
In Section 7 we find evidence that the stars comprising the `A' sections of the blue arm regions are older than the stars in the ring of active star formation, the NRC. This suggests a past epoch of star formation. 

The smoothness of the blue arms and their
intrinsic color suggest that their component stars are old enough to have dispersed from the cluster sites where they presumably formed. We have no information on the intrinsic (background corrected) color of the fainter `B' sections, because we can obtain no HST measures at these locations. Their slightly less blue color in Figure~\ref{Benedict.fig19} could be due entirely to the apparent greater spatial dispersion.  Our only resolved cluster, NRC 42, has a FWHM = 11.2pc. Its proximity to the blue arm (Figure 5), an age significantly older than clusters in the nuclear ring, and resolvable extent suggest cluster dispersion as an explanation for the smoothness of the blue arms.

Relative to these arms we identify two star formation rate (SFR) scenarios: 1) the blue arms could be due to a past epoch of high SFR further out than the present site of star formation. The rate in the blue arms is now essentially zero. The observed colors (less blue than the nuclear ring of active star formation) are then explained by the evolution of this population from very blue to less blue. 2) Alternately, the blue arms could be maintained by a present epoch of star formation. The observed colors are then explained by an intrinsically low SFR. Our results, namely no \Ha, no CO, and no young clusters local to the blue arms, all favor scenario 1.

%--------------------------------
%We have only weak evidence (the increased dispersion in the B sections of the blue arms seen in Figure~\ref{Benedict.fig19}) to support the contention that all the stars at the OILR are not the same age. In support of a uniform age hypothesis we point out that the IILR, the ellipse containing the NRC, has a semimajor axis of $a = 7\farcs4$. The ellipse containing the blue arms, identified with an OILR, has $a\sim$12\arcsec~(P1). Considering the OILR and IILR as triggering sites for star formation that activate as the peak of the gas distribution passes through them, the 2.1pc/Myr  rate of shrinking estimated above would predict past star formation at the locus of the OILR (the elliptical distribution containing the blue arms) $\sim$ 150 Myr ago. This agrees with our age estimate for the blue arms. The nuclear structure of NGC 4314 apparently depends on two processes. Inflow shrinks the reservoir of raw material required to fuel star formation (Combes et al. 1992). As the peak of that distribution of gas moved inward, gravitational instability associated with resonances (Elmegreen 1994) triggered star formation in the vicinity of the OILR, and now the IILR. This scenario was first explored for NGC 4321 through models and simulation by Knapen et al. (1995).
%-----------------------------

We have presented evidence for a past epoch of star formation exterior to the nuclear ring of present day star formation. This evidence supports the idea, first applied to NGC 4321 by Knapen et al. (1995) and summarized and extended by Shlosman (1996), that as gas inflows along the primary bar of a galaxy, star formation  occurs first near the OILR, then later near the IILR. Considering the OILR and IILR as triggering sites for star formation that activate as the bar-driven gas passes through them (Elmegreen 1994), the 2.1pc/Myr  rate of shrinking estimated above would predict past star formation at the locus of the OILR (the elliptical distribution containing the blue arms) $\sim$ 150 Myr ago. This agrees with our age estimate for the blue arms. Shlosman estimates that the entire process of migration through both the OILR and IILR takes less than a billion years. The nuclear structure of NGC 4314 apparently depends on two processes. Bar-driven inflow, first confirmed in P3,  compacts the reservoir of raw material required to fuel star formation (Combes et al. 1992). As that gas moved inward, gravitational instability associated with resonances (Elmegreen 1994) triggered past star formation at the OILR, in the vicinity of the blue arms,  and triggers it now at the IILR, the site of the present day nuclear ring.

\section{Summary}
	
We summarize the major observational results of this paper.
\begin{enumerate}

\item We have obtained UBVI\Ha~aperture photometry of 76 star clusters (NRC) associated with the nuclear ring of NGC 4314. 

\item We find that an ellipse can provide a good fit to the spatial distribution of NRC. If the NRC are coplanar with the principal plane of this galaxy and we assume a circular distribution, the ellipse fit yields an inclination, i=44\arcdeg, far greater than the i=23\arcdeg~consistent with the P1 outer isophotes. So, either the nuclear ring is intrinsically circular and tilted out of the plane of the outer disk or, alternatively, the distribution may be intrinsically elliptical (Ann 2001). If intrinsically elliptical, the NRC/blue arms could be an IILR/OILR, an identification supported by the CO dynamical results in P3 and the orientation difference of $\sim90\arcdeg$ (Combes 1988). 

\item We have applied the J\o rgensen et al. (1992) nested ellipse model to the nuclear region of NGC 4314. The Fourier coefficient, c$_4$, suggests a weakly boxy deviation from an elliptical distribution in the range 3 $<$ r $<$ 6 arcsec. We identify this effect with the bar hinted at in P2 and confirmed by Ann (2001).

\item By subtracting the smooth components of the galaxy light, we have mapped the dust (and, hence, gas) distribution interior to the nuclear ring, to within a few pc of the nucleus. The dust distribution is extremely complex, but some of the pattern is suggestive of those seen along the leading edges of bars in barred galaxies.  The dust distribution near the nuclear ring is strongly spiral.

\item From the reddening-free parameter, Q, and model cluster evolutionary tracks from SB99 we derived A$_V$ intrinsic to NGC 4314 for the NRC. The clusters in NE section of the nuclear ring have a characteristic A$_V\sim$ 1.5,
whereas those in the SW have a characteristic A$_V\sim$ 0.6,
suggesting that the NE side of the ring is the far side.

\item Corrected for interstellar extinction intrinsic to NGC 4314, the NRC colors are inconsistent with a single burst of star formation. The present episode of star formation has lasted at least $\sim$30 Myr.

\item The intercomparison of two reddening-free parameters (Q and Q$_3$) shows that the distribution of NRC ages peaks at $\sim$5 Myr. We find only weak evidence for either azimuthal or radial age gradients within the ring. It appears that star formation occurs in all parts of the ring nearly simultaneously. 

\item We have produced a continuum-corrected \Ha ~map of the nuclear ring region. From these data we obtain \Ha ~fluxes at the locations of a significant fraction (75\%) of the NRC. We find that \Ha ~correlates with our age parameter. We did not detect diffuse \Ha ~emission from the blue arm regions.

\item A comparison of a global SFR from total nuclear ring \Ha~flux with an SFR inferred from the duration of the present epoch of formation and cluster masses from UBVI photometry, suggests that a significant fraction of the star formation in the nuclear ring
occurs in the NRC clusters.

\item From photometry of residual flux in excess of that predicted by the nested ellipse model, we have determined that one past episode of star formation in NGC 4314 ended at least $50$ Myr ago. This is the minimum age for the stars in the blue arms and for the one star cluster (NRC 42) identified in the NW arm. Most of the locations in these arms have colors consistent with ages of 100-200 Myr.
We interpret the arms as the now-dispersed remnants of past clusters. 

\item Compared to NGC 1326 (\cite{But00}), the nuclear ring in NGC 4314 has an \Ha~luminosity (hence, SFR) and maximum cluster mass all a factor of ten less. Compared to NGC 1512 (Maoz \etal 2001) the nuclear ring in NGC 4314 has an \Ha~luminosity a factor of ten less, but a very similar cluster mass range. NGC 4314 has many clusters with directly associated \Ha~flux, unlike NGC 1326 and NGC 1512. NGC 4314 has half the internal extinction of NGC 1326 and about the same as NGC 1512. Neither NGC 1326 nor NGC 1512 have blue arms, sites of past recent star formation exterior to their nuclear rings. All three nuclear rings contain a similar small fraction of clusters older than 100 Myr.

\item The radii of the dynamically identified OILR and IILR, along with a rate of shrinkage inferred from the age range in the nuclear ring of active star formation, suggests that NGC 4314 presents us with a radially shrinking reservoir of raw material (molecular gas) that interacts with, in sequence, an OILR then an IILR. The age of the blue arms and the range of ages in the clusters associated with the IILR are consistent with this model.
\end{enumerate}

\acknowledgements 
Benedict thanks the {\it HST} Astrometry Science Team; W. Jefferys, P. Shelus, P. Hemenway, W. van Altena,
L. Fredrick, O. Franz, R. Duncombe, and B. McArthur for their support of this research that was far from astrometric in nature. Thanks also to Jim Jeletic, Ed Ruitberg, Dave Leckrone, Keith Kalinowski, and Ken Carpenter (all at Goddard Spaceflight Center) for their understanding, patient, and fair administration over these many years of the supporting NASA grant NAG5-1603. Benedict also thanks the Cox Endowment Fund of the University
of Texas Department of Astronomy
for sustaining support. We thank Jim Higdon for TIGER/OASIS \Ha~spectra in advance of publication. GFB thanks Ron Buta and Dani Maoz for discussions and a review of an ealier draft. Finally, to my patient friends in the HST cluster business, thank you for not grabbing these long-since public domain data.

The work by IJ on this research was supported by the Gemini Observatory, which is operated by the Association of
Universities for Research in Astronomy, Inc., under a cooperative agreement with the NSF on behalf
of the Gemini partnership: the National Science Foundation (United States), the Particle Physics and
Astronomy Research Council (United Kingdom), the National Research
Council (Canada), CONICYT (Chile), the Australian Research Council (Australia), CNPq (Brazil) and
CONICET (Argentina).

\clearpage

\begin{center}
\begin{deluxetable}{lrr}
\tablecaption{Global Properties of NGC 4314 \label{tbl-1}}
\tablewidth{0in}
\tablehead{\colhead{Parameter} &  \colhead{ Value }&  \colhead{ Reference }}
\startdata

Right ascension (J2000) & $12^h22^m32.0^s$& P1\\
Declination (J2000) &29\arcdeg 53' 44\farcs1 &P1\\
Classification &SBa& RC3\tablenotemark{a}\\
Inclination & 21\arcdeg & P3\\
Primary Bar Position angle & 158\arcdeg& P1\\
Systemic velocity, V$_{sys}$; CO &983$\pm$5 \kms & P3\\
Distance &13.1 Mpc\tablenotemark{b} & \\
Distance modulus& m-M = 30.59$\pm$0.15&\\
Linear Scale & 63.5 pc arcsec$^{-1}$&\\
Mass of molecular gas &$2.1 \times 10^8 {\cal M}_{\sun}$ & P3\\
\enddata
\tablenotetext{a} {Third Reference Catalog, \cite{GdV91}}
\tablenotetext{b} {Assuming H$_0$=75 km s$^{-1}$ Mpc$^{-1}$ (\cite{Mad99})}

\end{deluxetable}
\end{center}

\begin{center}
\begin{deluxetable}{lccrlrlrlrl}
\tablecaption{Nuclear Ring Cluster Photometry \label{tbl-2}}
\tablewidth{0in}
\tablehead{\colhead{NRC} &  \colhead{$\Delta$ RA\tablenotemark{a} }&  \colhead{$\Delta$ Dec\tablenotemark{a}}&  \colhead{V$_0$}&
&\colhead{(\ub)$_0$}& & \colhead{(\bv)$_0$}& &
\colhead{(\vi)$_0$} & }
\scriptsize
\startdata
1&5\farcs23&-9\farcs32&22.45&$\pm$0.05&-0.41&$\pm$0.13&0.22&$\pm$0.08&0.40&$\pm$0.09\\
2&1.24&-7.60&21.67&0.05&-0.62&0.08&-0.09&0.07&0.72&0.08\\
3&1.93&-7.39&22.04&0.06&-0.65&0.10&0.12&0.09&0.70&0.08\\
4&1.48&-6.73&21.62&0.07&-0.42&0.11&0.24&0.10&0.62&0.12\\
5&0.99&-6.63&21.35&0.05&-0.45&0.09&0.35&0.08&0.81&0.07\\
6&2.96&-6.23&21.55&0.06&-0.55&0.08&0.16&0.08&0.31&0.10\\
7&-0.34&-5.75&20.58&0.04&-0.82&0.05&-0.02&0.05&0.16&0.09\\
8&3.15&-5.70&20.04&0.02&-0.81&0.03&0.05&0.03&0.14&0.11\\
9&-0.04&-5.71&21.46&0.09&-0.70&0.14&0.21&0.13&0.93&0.13\\
10&-0.62&-5.44&19.46&0.02&-0.91&0.03&0.00&0.03&0.04&0.11\\
11&-0.69&-5.28&19.65&0.03&-0.91&0.04&0.07&0.05&0.11&0.10\\
12&-1.50&-5.20&21.50&0.07&-0.78&0.09&0.13&0.09&0.42&0.07\\
13&4.57&-5.06&21.98&0.06&-0.64&0.13&0.31&0.11&0.63&0.11\\
14&-0.87&-5.06&18.72&0.02&-0.87&0.02&-0.10&0.02&-0.11&0.04\\
15&-1.42&-5.09&21.24&0.06&-0.75&0.10&0.16&0.09&0.34&0.10\\
16&-2.16&-5.07&21.36&0.03&-0.86&0.05&-0.09&0.04&-0.39&0.12\\
17&4.19&-4.89&21.93&0.07&-0.60&0.13&0.42&0.12&0.55&0.08\\
18&-0.85&-4.77&19.21&0.03&-0.85&0.03&-0.05&0.03&-0.01&0.02\\
19&4.84&-4.71&21.59&0.05&-0.72&0.06&0.05&0.07&0.27&0.08\\
20&-1.28&-4.56&21.06&0.06&-0.93&0.05&-0.05&0.07&-0.19&0.03\\
21&-4.33&-3.89&21.80&0.04&-0.53&0.10&0.32&0.07&0.51&0.05\\
22&5.46&-3.72&21.08&0.04&-0.66&0.05&0.01&0.05&0.04&0.07\\
23&-4.36&-3.62&21.67&0.05&-0.75&0.08&0.14&0.08&0.10&0.06\\
24&5.67&-3.17&21.29&0.03&-0.59&0.07&0.57&0.06&0.44&0.07\\
25&-4.54&-2.69&21.19&0.06&-0.87&0.06&-0.08&0.08&0.18&0.10\\
\enddata
\tablenotetext{a} {Differential positions relative to galaxy center.}

\end{deluxetable}
\end{center}
\setcounter{table}{1}

\begin{center}
\begin{deluxetable}{lccrlrlrlrl}
\tablecaption{Nuclear Ring Cluster Photometry \label{tbl-2b}}
\tablewidth{0in}
\tablehead{\colhead{NRC} &  \colhead{$\Delta$ RA }&  \colhead{$\Delta$ Dec }&  \colhead{V$_0$}&
&\colhead{(\ub)$_0$}& & \colhead{(\bv)$_0$}& &
\colhead{(\vi)$_0$} & }
\scriptsize
\startdata
26&-4\farcs72&-2\farcs11&21.25&$\pm$0.08&-0.77&$\pm$0.08&0.05&$\pm$0.10&0.41&$\pm$0.13\\
27&-4.60&-1.74&20.66&0.03&-0.74&0.03&-0.02&0.03&0.58&0.04\\
28&6.60&-1.39&22.03&0.06&-0.42&0.12&0.18&0.08&0.83&0.08\\
29&6.43&-1.39&21.70&0.04&-0.49&0.09&0.34&0.06&0.69&0.04\\
30&5.67&-0.72&20.53&0.04&-0.78&0.04&0.00&0.05&0.04&0.13\\
31&5.49&-0.69&20.40&0.03&-0.80&0.05&0.13&0.05&0.25&0.09\\
32&-5.33&-0.63&22.28&0.06&-0.34&0.15&0.35&0.10&0.67&0.11\\
33&6.70&-0.31&19.80&0.01&-0.59&0.03&0.27&0.02&0.71&0.15\\
34&6.10&-0.34&19.63&0.03&-0.95&0.03&-0.15&0.03&-0.18&0.12\\
35&5.97&-0.27&19.94&0.02&-1.01&0.03&-0.08&0.03&-0.27&0.08\\
36&6.29&0.21&20.51&0.05&-0.90&0.05&-0.06&0.06&-0.19&0.06\\
37&-4.86&0.24&21.55&0.07&-0.32&0.09&0.17&0.09&0.61&0.12\\
38&-6.39&0.29&20.73&0.02&-0.80&0.03&0.08&0.03&-0.01&0.05\\
39&6.60&0.39&20.91&0.06&-0.77&0.09&0.10&0.08&0.44&0.08\\
40&6.60&0.66&21.13&0.04&-0.87&0.05&-0.03&0.05&0.25&0.05\\
41&4.43&1.44&22.05&0.05&-0.78&0.09&0.33&0.08&0.38&0.06\\
42&-8.95&2.04&22.81&0.07&-0.45&0.14&0.05&0.10&0.51&0.12\\
43&-6.32&2.13&21.29&0.08&-0.94&0.07&-0.09&0.09&0.16&0.13\\
44&-6.13&2.26&20.33&0.04&-0.88&0.04&-0.04&0.05&0.01&0.07\\
45&-6.35&2.30&21.14&0.06&-0.99&0.07&-0.04&0.08&-0.07&0.11\\
46&-6.31&2.71&20.86&0.05&-0.93&0.07&-0.07&0.07&0.24&0.09\\
47&5.43&2.89&21.06&0.03&-0.88&0.03&-0.07&0.04&-0.09&0.05\\
48&-5.59&3.13&20.62&0.07&-0.84&0.07&-0.08&0.08&-0.19&0.07\\
49&-5.93&3.16&20.29&0.03&-0.91&0.04&0.00&0.04&0.03&0.08\\
50&-5.67&3.29&20.23&0.04&-0.77&0.05&0.07&0.05&0.13&0.04\\
\enddata
\end{deluxetable}
\end{center}
\setcounter{table}{1}
\begin{center}
\begin{deluxetable}{lccrlrlrlrl}
\tablecaption{Nuclear Ring Cluster Photometry \label{tbl-2c}}
\tablewidth{0in}
\tablehead{\colhead{NRC} &  \colhead{$\Delta$ RA }&  \colhead{$\Delta$ Dec }&  \colhead{V$_0$}&
&\colhead{(\ub)$_0$}& & \colhead{(\bv)$_0$}& &
\colhead{(\vi)$_0$} & }
\scriptsize
\startdata
51&3\farcs53&3\farcs89&21.33&$\pm$0.05&-0.75&$\pm$0.07&0.15&$\pm$0.07&0.38&$\pm$0.09\\
52&-5.41&4.07&20.16&0.03&-0.53&0.05&0.36&0.05&0.48&0.11\\
53&2.50&4.11&21.04&0.05&-0.72&0.08&0.62&0.08&0.95&0.11\\
54&2.62&4.31&19.46&0.01&-0.84&0.02&0.06&0.02&0.26&0.05\\
55&-5.24&4.27&20.70&0.07&-0.72&0.11&0.22&0.11&0.23&0.09\\
56&-5.11&4.33&20.66&0.07&-0.63&0.07&0.10&0.09&0.32&0.10\\
57&2.55&4.42&19.40&0.01&-0.87&0.02&0.10&0.02&0.22&0.05\\
58&-5.79&4.72&21.10&0.04&-0.94&0.05&-0.08&0.05&-0.12&0.07\\
59&-5.34&4.93&21.03&0.06&-0.75&0.10&0.00&0.09&0.38&0.03\\
60&-5.23&4.93&20.75&0.04&-0.71&0.07&-0.04&0.06&0.37&0.08\\
61&-4.88&5.31&21.12&0.07&-0.76&0.11&0.07&0.10&0.20&0.08\\
62&1.89&5.48&20.72&0.02&-0.51&0.06&0.45&0.04&0.56&0.07\\
63&1.85&5.68&21.30&0.05&-0.59&0.08&0.32&0.07&0.35&0.11\\
64&-3.27&5.91&20.48&0.03&-0.66&0.04&0.19&0.04&0.58&0.05\\
65&1.21&5.84&22.11&0.07&-0.56&0.14&0.13&0.11&0.50&0.08\\
66&-0.47&5.88&20.95&0.05&-0.78&0.07&0.29&0.08&0.33&0.06\\
67&-3.30&5.90&20.51&0.03&-0.69&0.05&0.18&0.05&0.58&0.07\\
68&-4.21&5.89&20.96&0.06&-0.74&0.08&0.00&0.08&0.30&0.09\\
69&-4.17&5.99&20.40&0.03&-0.80&0.05&0.03&0.05&0.14&0.07\\
70&-3.31&6.05&20.66&0.06&-0.63&0.07&0.09&0.08&0.35&0.09\\
71&-3.43&6.12&20.99&0.06&-0.60&0.09&0.10&0.09&0.30&0.12\\
72&-4.21&6.18&20.44&0.03&-0.82&0.05&0.00&0.04&0.04&0.06\\
73&-0.42&6.49&21.79&0.05&-0.52&0.14&0.50&0.10&0.79&0.07\\
74&0.46&6.77&21.94&0.07&-0.79&0.09&-0.03&0.09&0.27&0.11\\
75&-0.27&7.30&21.71&0.04&-0.76&0.07&0.02&0.06&0.42&0.07\\
76&-1.91&7.51&21.46&0.03&0.00&0.15&0.55&0.07&0.60&0.05\\
\enddata
\end{deluxetable}
\end{center}

\clearpage

\begin{center}
\begin{deluxetable}{lccrlrlrlrl}
\tablecaption{Surface Photometry of Locations within Blue Arms \label{tbl-3}}
\tablewidth{0in}
\tablehead{\colhead{ID} &  \colhead{$\Delta$ RA\tablenotemark{a}  }&  \colhead{$\Delta$ Dec\tablenotemark{a} }&  \colhead{$\mu_{V_0}$}&
&\colhead{$\mu_{(U-B)_0}$} & & \colhead{$\mu_{(B-V)_0}$} & &
\colhead{$\mu_{(V-I)_0}$} & }
\scriptsize
\startdata
A1&9\farcs20&-2\farcs87&21.97&$\pm$0.04&-0.16&$\pm$0.10&0.15&$\pm$0.05&0.49&$\pm$0.05\\
A2&8.88&-3.73&21.99&0.04&-0.14&0.10&0.17&0.05&0.49&0.05\\
A3&8.33&-4.51&21.73&0.03&-0.27&0.08&0.31&0.05&0.46&0.04\\
A4&7.74&-5.15&21.62&0.03&-0.10&0.08&0.15&0.04&0.56&0.04\\
A5&6.83&-5.78&21.74&0.03&-0.25&0.08&0.28&0.05&0.67&0.04\\
A6&6.01&-6.15&21.74&0.04&-0.35&0.08&0.39&0.05&0.73&0.04\\
A7&-9.24&2.78&22.12&0.04&-0.25&0.10&0.20&0.06&0.24&0.06\\
A8&-8.70&3.92&21.80&0.03&-0.14&0.09&0.13&0.05&0.48&0.04\\
A9&-8.02&4.78&21.91&0.04&-0.11&0.11&0.30&0.05&0.53&0.05\\
A10&-7.47&5.28&22.13&0.05&-0.23&0.12&0.38&0.06&0.56&0.05\\
A11&-6.60&6.10&21.93&0.04&-0.02&0.12&0.28&0.05&0.51&0.05\\
A12&-8.93&2.00&22.12&0.04&-0.23&0.08&-0.17&0.05&0.69&0.05\\
\enddata
\tablenotetext{a} {Differential positions relative to galaxy center.}

\end{deluxetable}
\end{center}
\setcounter{table}{3}
\clearpage
\begin{center}
\begin{deluxetable}{crlrlc}
\tablecaption{Nuclear Ring Cluster Derived Quantities - A$_V$, M$_V$, and ${\cal M}$ \label{tbl-4}}
\tablewidth{0in}
\tablehead{\colhead{NRC} &  \colhead{ A$_V$ }& & \colhead{M$_V$}& & \colhead{${\cal M}_{NRC}$ }\\
 &  & & & & \colhead{(10$^4{\cal M}_{\sun}$})}
\scriptsize
\startdata
1&0.78&0.26&-8.92&$\pm$0.30&0.38\\
2&0.00&0.07&-8.92&0.14&1.78\\
3&0.52&0.28&-9.07&0.30&1.16\\
4&0.84&0.31&-9.81&0.36&1.40\\
5&1.20&0.23&-10.44&0.25&2.77\\
6&0.61&0.25&-9.66&0.30&0.72\\
7&0.15&0.15&-10.16&0.19&1.22\\
8&0.41&0.08&-10.97&0.10&1.85\\
9&0.90&0.40&-10.03&0.42&3.22\\
10&0.34&0.08&-11.48&0.09&2.42\\
11&0.61&0.14&-11.55&0.16&4.61\\
12&0.67&0.28&-9.76&0.33&0.92\\
13&1.22&0.33&-9.83&0.38&0.95\\
14&0.00&0.02&-11.87&0.05&3.85\\
15&0.78&0.28&-10.13&0.33&0.97\\
16&0.00&0.04&-9.23&0.14&0.18\\
17&1.62&0.37&-10.28&0.41&0.79\\
18&0.03&0.10&-11.42&0.12&2.95\\
19&0.30&0.21&-9.30&0.25&0.64\\
20&0.16&0.22&-9.70&0.34&0.34\\
21&1.14&0.22&-9.93&0.24&0.89\\
22&0.14&0.16&-9.65&0.21&0.59\\
23&0.67&0.24&-9.59&0.33&0.37\\
24&2.26&0.18&-11.56&0.23&0.99\\
25&0.00&0.08&-9.40&0.19&0.77\\
\enddata
\end{deluxetable}
\end{center}
\setcounter{table}{3}
\clearpage
\begin{center}
\begin{deluxetable}{crlrlc}
\tablecaption{Nuclear Ring Cluster Derived Quantities - A$_V$, M$_V$, and ${\cal M}$ \label{tbl-4b}}
\tablewidth{0in}
\tablehead{\colhead{NRC} &  \colhead{ A$_V$ }& & \colhead{M$_V$}& & \colhead{${\cal M}_{NRC}$ }\\
 &  & & & & \colhead{(10$^4{\cal M}_{\sun}$})}
\scriptsize
\startdata
26&0.37&0.31&-9.71&$\pm$0.38&1.16\\
27&0.08&0.11&-10.01&0.12&3.15\\
28&0.65&0.25&-9.21&0.27&1.59\\
29&1.18&0.20&-10.08&0.22&1.52\\
30&0.17&0.15&-10.23&0.19&0.97\\
31&0.71&0.15&-10.90&0.17&1.65\\
32&1.17&0.31&-9.49&0.36&0.85\\
33&0.16&0.07&-10.95&0.07&4.06\\
34&0.00&0.03&-10.96&0.09&1.29\\
35&0.16&0.09&-10.82&0.13&0.74\\
36&0.08&0.20&-10.17&0.26&0.57\\
37&0.69&0.29&-9.73&0.35&3.15\\
38&0.52&0.10&-10.38&0.12&0.70\\
39&0.56&0.25&-10.25&0.28&1.67\\
40&0.16&0.17&-9.62&0.23&0.89\\
41&1.56&0.25&-10.10&0.33&0.44\\
42&0.27&0.31&-8.06&0.37&0.60\\
43&0.02&0.29&-9.32&0.35&0.62\\
44&0.13&0.15&-10.39&0.18&1.04\\
45&0.29&0.26&-9.75&0.32&0.39\\
46&0.05&0.22&-9.78&0.24&1.11\\
47&0.00&0.11&-9.54&0.15&0.44\\
48&0.00&0.08&-9.97&0.25&0.57\\
49&0.34&0.13&-10.64&0.15&1.09\\
50&0.45&0.16&-10.81&0.19&1.53\\
\enddata
\end{deluxetable}
\end{center}
\setcounter{table}{3}
\clearpage
\begin{center}
\begin{deluxetable}{crlrlc}
\tablecaption{Nuclear Ring Cluster Derived Quantities - A$_V$, M$_V$, and ${\cal M}$ \label{tbl-4c}}
\tablewidth{0in}
\tablehead{\colhead{NRC} &  \colhead{ A$_V$ }& & \colhead{M$_V$}& & \colhead{${\cal M}_{NRC}$ }\\
 &  & & & & \colhead{(10$^4{\cal M}_{\sun}$})}
\scriptsize
\startdata
51&0.70&0.21&-9.96&$\pm$0.26&0.98\\
52&1.29&0.15&-11.72&0.17&3.69\\
53&2.20&0.25&-11.75&0.28&4.23\\
54&0.49&0.06&-11.63&0.07&4.14\\
55&0.98&0.34&-10.87&0.40&1.21\\
56&0.42&0.29&-10.35&0.34&1.69\\
57&0.69&0.05&-11.88&0.06&3.80\\
58&0.03&0.17&-9.52&0.22&0.38\\
59&0.15&0.27&-9.71&0.30&1.39\\
60&-0.01&0.20&-9.83&0.22&1.78\\
61&0.40&0.31&-9.87&0.36&0.81\\
62&1.62&0.13&-11.49&0.14&2.57\\
63&1.18&0.21&-10.48&0.25&0.94\\
64&0.77&0.12&-10.88&0.14&3.59\\
65&0.51&0.34&-9.00&0.40&0.67\\
66&1.36&0.24&-11.00&0.28&1.11\\
67&0.75&0.15&-10.84&0.17&3.40\\
68&0.15&0.26&-9.78&0.29&1.21\\
69&0.29&0.14&-10.49&0.17&1.34\\
70&0.38&0.24&-10.31&0.28&1.83\\
71&0.43&0.27&-10.03&0.31&1.20\\
72&0.23&0.14&-10.38&0.17&1.02\\
73&1.84&0.31&-10.64&0.32&1.63\\
74&0.07&0.28&-8.72&0.33&0.46\\
75&0.22&0.20&-9.11&0.23&0.80\\
76&1.90&0.21&-11.03&0.22&3.67\\
\enddata
\end{deluxetable}
\end{center}

\clearpage
\begin{center}
\begin{deluxetable}{lrlcrc}
\tablecaption{\Ha~Associated with Nuclear Ring Clusters  \label{tbl-5a}}
\tablewidth{0in}
\tablehead{\colhead{NRC} &  \colhead{ \Ha~flux }& & \colhead{L$_H\alpha$}& \colhead{ m$^{c}_{H\alpha}$}  & \colhead{Age Flag \tablenotemark{a}}\\
 & \colhead{(10$^{-15}$)} & &(log) & &  }
\scriptsize
\startdata
1&0.11&$\pm$0.11&36.3&21.2&O\\
2&0.05&0.11&36.0&22.7&O\\
3&0.12&0.11&36.4&21.3&O\\
4&0.37&0.12&36.9&19.9&O\\
5&0.39&0.12&36.9&19.5&O\\
6&0.44&0.12&37.0&19.8&O\\
7&1.04&0.13&37.3&19.3&Y\\
8&0.14&0.11&36.5&21.2&Y\\
9&1.80&0.15&37.6&18.1&Y\\
10&0.22&0.11&36.7&20.8&Y\\
11&0.55&0.12&37.1&20.0&Y\\
12&1.55&0.15&37.5&18.4&Y\\
13&0.46&0.12&37.0&19.3&Y\\
14&0.17&0.11&36.6&21.4&Y\\
15&1.63&0.15&37.5&18.3&Y\\
16&0.73&0.13&37.2&19.8&Y\\
17&1.47&0.14&37.5&17.7&Y\\
18&0.77&0.13&37.2&19.7&Y\\
19&0.59&0.12&37.1&19.8&Y\\
20&1.32&0.14&37.4&19.0&Y\\
21&0.87&0.13&37.3&18.7&Y\\
22&1.04&0.14&37.3&19.3&O\\
23&1.50&0.14&37.5&18.5&Y\\
24&0.45&0.12&37.0&18.5&Y\\
25&0.45&0.12&37.0&20.4&Y\\
\enddata
\tablenotetext{a} { Age $<$ 15 Myr = Y}
\end{deluxetable}
\end{center}

\setcounter{table}{4}
\clearpage
\begin{center}
\begin{deluxetable}{lrlcrc}
\tablecaption{\Ha~Associated with Nuclear Ring Clusters  \label{tbl-5b}}
\tablewidth{0in}
\tablehead{\colhead{NRC} &  \colhead{ \Ha~flux }& & \colhead{L$_H\alpha$}& \colhead{ m$^{c}_{H\alpha}$}  & \colhead{Age Flag}\\
 & \colhead{(10$^{-15}$)} & &(log) & &  }
\scriptsize
\startdata
26&0.41&$\pm$0.12&36.9&20.1&Y\\
27&0.28&0.11&36.8&20.7&O\\
28&0.65&0.12&37.1&19.4&O\\
29&0.77&0.13&37.2&18.8&O\\
30&0.98&0.13&37.3&19.3&Y\\
31&0.92&0.13&37.3&19.0&Y\\
32&0.49&0.12&37.0&19.3&O\\
33&1.09&0.14&37.4&18.5&Y\\
34&1.84&0.15&37.6&18.8&Y\\
35&2.09&0.16&37.6&18.5&Y\\
36&1.19&0.14&37.4&19.2&Y\\
37&0.33&0.11&36.8&20.5&O\\
38&0.52&0.12&37.0&19.7&Y\\
39&1.13&0.14&37.4&18.9&Y\\
40&1.35&0.14&37.4&19.0&Y\\
41&0.78&0.13&37.2&18.5&Y\\
42&0.12&0.11&36.4&21.8&O\\
43&0.46&0.12&37.0&20.3&Y\\
44&0.57&0.12&37.1&20.0&Y\\
45&0.32&0.11&36.8&20.5&Y\\
46&0.32&0.11&36.8&20.6&Y\\
47&2.06&0.16&37.6&18.7&Y\\
48&1.32&0.14&37.4&19.2&Y\\
49&1.91&0.16&37.6&18.5&Y\\
50&2.20&0.16&37.7&18.2&Y\\
\enddata
\end{deluxetable}
\end{center}

\setcounter{table}{4}
\clearpage
\begin{center}
\begin{deluxetable}{lrlcrc}
\tablecaption{\Ha~Associated with Nuclear Ring Clusters  \label{tbl-5c}}
\tablewidth{0in}
\tablehead{\colhead{NRC} &  \colhead{ \Ha~flux }& & \colhead{L$_H\alpha$}& \colhead{ m$^{c}_{H\alpha}$} & \colhead{Age Flag}\\
 & \colhead{(10$^{-15}$)} & &(log) & &  }
\scriptsize
\startdata
51&0.52&$\pm$0.12&37.0&19.6&Y\\
52&0.83&0.13&37.2&18.6&Y\\
53&2.21&0.16&37.7&16.8&Y\\
54&11.80&0.31&38.4&16.4&Y\\
55&1.06&0.13&37.3&18.6&Y\\
56&1.07&0.13&37.3&19.0&O\\
57&13.90&0.33&38.5&16.0&Y\\
58&0.54&0.12&37.0&20.1&Y\\
59&0.47&0.12&37.0&20.1&O\\
60&0.64&0.12&37.1&19.9&O\\
61&0.47&0.12&37.0&20.0&Y\\
62&3.32&0.19&37.8&16.8&Y\\
63&2.08&0.16&37.6&17.7&Y\\
64&1.11&0.14&37.4&18.7&Y\\
65&0.97&0.13&37.3&19.1&O\\
66&0.98&0.13&37.3&18.4&Y\\
67&1.15&0.14&37.4&18.7&Y\\
68&0.22&0.11&36.7&21.0&O\\
69&0.09&0.11&36.3&21.8&Y\\
70&0.89&0.13&37.3&19.3&O\\
71&0.74&0.13&37.2&19.4&O\\
72&0.11&0.11&36.4&21.6&Y\\
73&0.21&0.11&36.6&19.7&Y\\
74&0.24&0.11&36.7&20.9&Y\\
75&0.10&0.11&36.3&21.8&Y\\
76&&no&detection& & \\
\enddata
\end{deluxetable}
\end{center}

\clearpage

\clearpage

\clearpage

\clearpage

\clearpage

\clearpage

\clearpage

\clearpage

\clearpage

\clearpage

\clearpage

\clearpage

\clearpage

\clearpage

\clearpage

\clearpage

\clearpage

\clearpage

\clearpage

\end{document}